\documentclass[prd,preprint,superscriptaddress,tightenlines,nofootinbib,
  eqsecnum,showpacs]{revtex4}

\usepackage{amsmath}
\usepackage{amsfonts}
\usepackage{amssymb}
\usepackage{bm}
\usepackage{hyperref}
\usepackage{mathrsfs}
\usepackage{graphicx}

\usepackage{ulem}
\usepackage[usenames]{color}

\definecolor{darkgreen}{rgb}{0,0.5,0}

\hypersetup{
    bookmarks=true,         
    unicode=false,          
    pdftoolbar=true,        
    pdfmenubar=true,        
    pdffitwindow=false,     
    pdfstartview={FitH},    
    pdftitle={My title},    
    pdfauthor={Author},     
    pdfsubject={Subject},   
    pdfcreator={Creator},   
    pdfproducer={Producer}, 
    pdfkeywords={keyword1} {key2} {key3}, 
    pdfnewwindow=true,      
    colorlinks=true,       
    linkcolor=red,          
    citecolor=cyan,        
    filecolor=magenta,      
    urlcolor=darkgreen,           
    linktocpage=true
}

\allowdisplaybreaks

\DeclareSymbolFontAlphabet{\mathrsfs}{rsfs}
\DeclareMathAlphabet{\mathcal}{OMS}{cmsy}{m}{n}

\newcommand{\ud}{\mathrm{d}}
\newcommand{\ui}{\mathrm{i}} 
\newcommand{\ue}{\mathrm{e}} 
\newcommand{\beq}{\begin{equation}}
\newcommand{\eeq}{\end{equation}}

\newcounter{theorem} 

\begin{document}

\title{Dimensional regularization of the IR divergences \\in the
  Fokker action of point-particle binaries \\at the fourth
  post-Newtonian order}

\author{Laura Bernard}\email{laura.bernard@tecnico.ulisboa.pt}
\affiliation{CENTRA, Departamento de F\'{\i}sica, Instituto Superior
  T{\'e}cnico -- IST, Universidade de Lisboa -- UL, Avenida Rovisco
  Pais 1, 1049 Lisboa, Portugal}

\author{Luc Blanchet}\email{blanchet@iap.fr}
\affiliation{$\mathcal{G}\mathbb{R}\varepsilon{\mathbb{C}}\mathcal{O}$,
  Institut d'Astrophysique de Paris --- UMR 7095 du CNRS,
  \\ Universit\'e Pierre \& Marie Curie, 98\textsuperscript{bis}
  boulevard Arago, 75014 Paris, France}

\author{Alejandro Boh\'e}\email{alejandro.bohe@aei.mpg.de}
\affiliation{Max Planck Institute for Gravitational Physics (Albert Einstein
  Institute), Am Muehlenberg 1, 14476
  Potsdam-Golm, Germany}

\author{Guillaume Faye}\email{faye@iap.fr}
\affiliation{$\mathcal{G}\mathbb{R}\varepsilon{\mathbb{C}}\mathcal{O}$,
  Institut d'Astrophysique de Paris --- UMR 7095 du CNRS,
  \\ Universit\'e Pierre \& Marie Curie, 98\textsuperscript{bis}
  boulevard Arago, 75014 Paris, France}

\author{Sylvain Marsat}\email{sylvain.marsat@aei.mpg.de}
\affiliation{Max Planck Institute for Gravitational Physics (Albert Einstein
  Institute), Am Muehlenberg 1, 14476
  Potsdam-Golm, Germany} 

\date{\today}

\begin{abstract}
The Fokker action of point-particle binaries at the fourth
post-Newtonian (4PN) approximation of general relativity has been
determined previously. However two ambiguity parameters associated
with infra-red (IR) divergencies of spatial integrals had to be
introduced. These two parameters were fixed by comparison with
gravitational self-force (GSF) calculations of the conserved energy
and periastron advance for circular orbits in the test-mass limit. In
the present paper together with a companion paper, we determine both these ambiguities from first
principle, by means of dimensional regularization. Our computation is
thus entirely defined within the dimensional regularization scheme,
for treating at once the IR and ultra-violet (UV) divergencies. In
particular, we obtain crucial contributions coming from the Einstein-Hilbert part of the action and from the non-local
tail term in arbitrary dimensions, which resolve the ambiguities.
\end{abstract}

\pacs{04.25.Nx, 04.30.-w, 97.60.Jd, 97.60.Lf}

\maketitle

\section{Introduction} 
\label{sec:intro}


In previous works~\cite{BBBFMa,BBBFMb} (respectively referred to as
Papers~I and~II), we determined the Fokker Lagrangian of the motion of
compact binary systems (without spins) in harmonic coordinates at the
fourth post-Newtonian (4PN) approximation of general
relativity.\footnote{As usual the $n$PN order means the terms of order
  $(v/c)^{2n}$ in the equations of motion relatively to the Newtonian
  acceleration.} Equivalent results had been previously achieved using
the ADM Hamiltonian formalism, in ADM-like coordinates, developed at
4PN order~\cite{JaraS12,JaraS13,JaraS15,DJS14,DJS16}. Partial results
have been obtained at the 4PN order using the effective field theory
(EFT) approach~\cite{FS4PN,FStail,GLPR16,FMSS16}. A prominent feature
of this order is the non-locality (in time) due to the imprint of
gravitational wave tails starting at that approximation (see
also~\cite{BL17}).

We start with the gravitation-plus-matter action, made of the gauged-fixed
Einstein-Hilbert action of general relativity plus the matter terms describing
point particles. The Fokker action governing the motion of compact binaries is
then obtained by replacing the generic metric in the complete action by an
explicit post-Newtonian solution of the corresponding Einstein field
equations. The PN metric is parametrized by appropriate PN potentials, which
are obtained as explicit functionals of the particles' parameters and
trajectories. The maximal PN order to which each of the components of the
metric is to be controlled and inserted into the action is determined by the
method called ``$n+2$'' in Paper~I (see Sec.~IV A there). The spatial
integrals coming from the Einstein-Hilbert part of the action are computed in
the physical domain, using elementary solutions of the Poisson equation (the
Fock kernel~\cite{Fock} as well as its generalizations) and, in a first stage, the
Hadamard ``partie finie'' integral in $3$ dimensions, which is equivalent to a
Riesz integration~\cite{Riesz}.

In a second stage, as will be reported in the present paper, we correct the
calculation so as to take into account dimensional regularization and the
presence of poles in the dimension, in the limit where $d-3\to 0$, for both UV
and IR type divergences. Finally, as the non-local tail term is not included
into the ``$n+2$'' method, we have to compute it in $d$ dimensions and add it
separately to the action. The resulting Fokker Lagrangian is a generalized
one, depending on accelerations and derivatives of accelerations. We reduce it
to a simpler Lagrangian linear in accelerations (in harmonic coordinates) by
adding suitable multi-zero terms and total time derivatives.

Carefully choosing and implementing regularizations play a crucial role in
this field. In Papers~I and~II, we adopted a dimensional regularization scheme
for treating the ultra-violet (UV) divergences associated with
point-particles, as well as a Hadamard regularization for curing the infra-red
(IR) divergences occurring at the bound at infinity of integrals in the
gravitational part of the Fokker action (as we know, IR divergences start
occurring precisely at the 4PN order). Unfortunately, we had to introduce in
Paper~I an ``\textit{ambiguity parameter}'' reflecting some incompleteness in
the Hadamard treatment of the IR divergences. This ambiguity was then fixed by
matching the conserved energy in the case of circular orbits to known results
obtained from gravitational self-force (GSF) calculations in the test-mass
limit~\cite{BDLW10b,LBW12,LBB12,BiniD13}. Note that an equivalent ambiguity
parameter had also to be included in the ADM Hamiltonian
formalism~\cite{DJS14}. Furthermore, we were forced to add in Paper~II a
\textit{second} ambiguity parameter in order to match the periastron advance
for circular orbits with the results coming from GSF calculations. The latter
results are known from numerical~\cite{BDS10,Letal11,vdM16}
and analytical~\cite{D10sf,DJS15eob,DJS16,BL17} studies. As we conjectured in
Paper~II, this second ambiguity parameter was in fact mandatory, since the
difference between different prescriptions for the IR regularization of
integrals at infinity can be reduced, after a suitable shift of the
world-lines, to two and only two offending terms at the 4PN order in the
Lagrangian.

The aim of the present paper and of the companion paper~\cite{MBBF17} is to
resolve the issue of the two ambiguity parameters, \textit{i.e.}, to compute
their values from \textit{first principles}.\footnote{In a first version of
  the present paper, due to an incomplete implementation of the regularization
  procedure for the matching between near and far zones in the computation of
  the tail term, we could only solve for the second ambiguity parameter. In
  the companion paper~\cite{MBBF17}, we carefully implement this
  regularization, and show that it yields the correct value of the first
  ambiguity parameter.} To do so, we employ the powerful dimensional
regularization~\cite{tHooft,Bollini,Breitenlohner} (instead of Hadamard's) for
resolving the IR divergences of the Fokker action occurring at the bound at
infinity of spatial integrals. Therefore, our Fokker action will now be
entirely based on dimensional regularization, for both the IR and UV
divergences. We have two main tasks:
\begin{enumerate}
\item Computing the \textit{difference} between the dimensional regularized
  and the Hadamard regularized gravitational (\textit{i.e.}, Einstein-Hilbert)
  parts of the Fokker action. For this calculation we shall use known formulas
  for the ``difference'' between these two regularizations coming from
  Refs.~\cite{BDE04,BDEI05dr}. The needed accuracy of the post-Newtonian
  calculation will follow the rules of the method $n+2$ in Paper~I;
  
\item Evaluating the non-local tail term in $d$ dimensions or, rather, an
  associated homogeneous solution that is to be added to the ``difference''
  computed from the $n+2$ method. The precise way in which the 4PN tail effect
  enters our calculation is through the ``matching'' equation, whose solution
  gives a connection between the near zone and the far zone where tails
  propagate. This equation is the key to the final completion of the problem
  and the computation of the ambiguity parameters.\footnote{We refer to the
    companion paper~\cite{MBBF17} for more details about the matching equation
    and the overall calculation.} We find that the calculation reduces to that
  of a series of elementary non-local integrals, multiplied by some non
  trivial numerical coefficient, which is computed in closed analytic form
  with Euler gamma functions in App.~\ref{app:coeffC}. As we shall see (and in
  agreement with EFT works~\cite{FStail,GLPR16,PR17,Po17}), such a
  tail-induced homogeneous solution contains a UV-like pole in $d$ dimensions.
  We shall prove that this pole precisely cancels the IR-like pole remaining
  from the $n+2$ method after applying suitable shifts, while the finite part
  gives a suplementary contribution of the form of the ambiguity parameters of
  Paper~II.
\end{enumerate}
Adding up the contributions from the latter two steps (and also, subtracting
off a particular surface term in our previous Hadamard IR regularization
scheme), we finally find that the modification of the Lagrangian takes exactly
the form postulated in Paper~II. Moreover, we find that the two ambiguity
parameters $\delta_1$ and $\delta_2$ (following exactly the definition in
Sec.~II of Paper~II) are in complete agreement with the result of Paper~II
[see Eq.~(2.6) there], so that the corresponding conserved energy and
periastron advance for circular orbits at 4PN order are correct. We conclude
that our 4PN dynamics based on the Fokker action in harmonic coordinates is
now complete.

We see that the calculation crucially relies on dimensional regularization and
one may wonder why this regularization finally gives the correct answer. We
are in fact borrowing this technique to quantum field theory and EFT~\cite{GR06},
since dimensional regularization was invented as a mean to preserve the gauge
invariance of quantum gauge field
theories~\cite{tHooft,Bollini,Breitenlohner}. In the present context,
dimensional regularization serves at preserving the diffeomorphism invariance
of general relativity. It permits to respect the basic properties of
algebraic and differential calculations, such as the associativity,
commutativity and distributivity of point-wise addition and multiplication,
the Leibniz and Schwarz rules, and the integration by parts~\cite{DJSdim}. We
argue that, for this reason, dimensional regularization is the only known mean
to obtain directly the correct answer to the problem of self interacting point
masses at the 4PN order.

The plan of this paper is as follows. In Sec.~\ref{sec:diffIR} we obtain the
difference between the dimensional and Hadamard IR regularizations for the
gravitational part of the Fokker action. After application of shifts we find
that such a difference contains a residual IR pole. In Sec.~\ref{sec:NZ}, we
investigate general technical formulas for the computation of the near zone
expansion of the solution of the wave equation in $d$ dimensions. These
formulas are then applied in Sec.~\ref{sec:tail} to the derivation of the tail
term in the near zone metric and then in the Fokker action. We obtain a UV
pole that exactly cancels the IR pole coming from the gravitational part of
the Fokker action. This determines the second ambiguity (Sec.~\ref{sec:amb}).
Moreover, thanks to a careful matching between the near zone and the far zone in our
formalism, we are able to determine the first ambiguity parameter as well, as
shown in~\cite{MBBF17}. Technical appendices provide important material on:
the homogeneous solutions of the wave equation and their PN expansion in
App.~\ref{app:hom}; the multipole expansion of elementary functions and
potentials in $d$ dimensions in App.~\ref{app:mult}; some distributional
limits of Green's functions in App.~\ref{app:distr}; the computation of some
particular intricate coefficient in App.~\ref{app:coeffC}.

\section{Dimensional regularization of infra-red divergences}
\label{sec:diffIR}

In Paper~I~\cite{BBBFMa} it was shown that IR divergences, due to the
behaviour of spatial integrals at infinity, start to appear at the 4PN
order in the Fokker action of general matter systems. These IR
divergences are associated with non-local tail effects in the dynamics
occuring at 4PN order~\cite{BD88,B93}. In Paper~I it was found that
two arbitrary scales respectively associated with tails (denoted $s_0$
in Paper~I) and the IR cut-off (denoted $r_0$) combine to give an
``ambiguity'' parameter $\alpha=\ln(r_0/s_0)$ which could not be
determined within the method. Equivalent results had been obtained
with the Hamiltonian formalism in Ref.~\cite{DJS14}. However, in
contrast to the Hamiltonian formalism, we had to introduce in Paper~II
a second ambiguity parameter and argued that it was due to our
particular treatment of the IR divergences based on the Hadamard
``partie finie'' integral. On the other hand, the UV divergences
associated with point particles were cured by dimensional
regularization.

In the present paper we shall employ dimensional regularization for
both the IR and UV divergences. As we shall see, using dimensional
regularization does modify the end result for the Fokker Lagrangian
(and associated Hamiltonian), but in a way that is fully consistent
with the conjecture put forward in Paper~II. Therefore this justifies
the final 4PN dynamics obtained in Paper~II and in particular, we
confirm that the 4PN dynamics is compatible with existing GSF
computations of the energy and periastron advance for circular orbits.

We want to regularize the three-dimensional divergent integral
\begin{equation}\label{I}
I = \int \ud^3 \mathbf{x}\,F(\mathbf{x})\,,
\end{equation}
where the function $F$ is obtained by following the PN iteration
procedure of the field equations using the method $n+2$ (see Sec.~IV A
of Paper~I). The integral~\eqref{I} represents a generic term in the
gravitational (Einstein-Hilbert) part of the Fokker Lagrangian
$L_g$. Specifically, since we are dealing with the IR bound at
infinity, we consider
\begin{equation}\label{IcalR}
I_\mathcal{R} = \int_{r > \mathcal{R}}\ud^3
\mathbf{x}\,F(\mathbf{x})\,,
\end{equation}
where the integration domain is restricted to be
$r=\vert\mathbf{x}\vert>\mathcal{R}$, with $\mathcal{R}$ being a
sufficiently large constant radius. The divergences occur from the
expansion of $F$ when $r\to +\infty$, which is of the type (for any
$N\in\mathbb{N}$)
\begin{equation}\label{Fdev0}
F(\mathbf{x}) = \sum_{p=-p_0}^{N}\frac{1}{r^p}\,\varphi_p(\mathbf{n})
+ o\left(\frac{1}{r^N}\right)\,.
\end{equation}
The coefficients $\varphi_p$ depend on the unit direction
$\mathbf{n}=\mathbf{x}/r$ and on $p\in\mathbb{Z}$; the minimal value
of $p$ corresponds to some highly divergent behaviour with growing
power $\sim r^{p_0}$ of the distance. In what follows we shall write
for simplicity some formal expansion series without expliciting the
remainder term, that is
\begin{equation}\label{Fdev}
F(\mathbf{x}) = \sum_{p\geqslant
  -p_0}\frac{1}{r^p}\,\varphi_p(\mathbf{n})\,.
\end{equation}

In Paper~I a regularization factor $(r/r_0)^B$ was introduced into the
integrand and the integral was considered in the sense of analytic
continuation in $B\in\mathbb{C}$. Then the regularized value of the
integral was defined as the finite part (FP), \textit{i.e.}, the
coefficient of the zero-th power of $B$, in the Laurent expansion of
the regularized integral when $B\to 0$. This prescription, which is
equivalent to a Hadamard regularization (HR), reads
\begin{equation}\label{IHRdef}
I_\mathcal{R}^\text{HR} = \mathop{\text{{\rm
      FP}}}_{B=0}\int_{r>\mathcal{R}} \ud^{3}\mathbf{x}
\,\Bigl(\frac{r}{r_0}\Bigr)^B F(\mathbf{x}) \,.
\end{equation}
A straightforward calculation, plugging~\eqref{Fdev}
into~\eqref{IcalR} (where $\mathcal{R}$ is a large radius), yields the
HR-regularized version of the integral as
\begin{equation}\label{IHR}
I_\mathcal{R}^\text{HR} = - \sum_{p\not=
  3}\frac{\mathcal{R}^{3-p}}{3-p}\,\int\ud\Omega_2\,\varphi_p(\mathbf{n})
- \ln\left(\frac{\mathcal{R}}{r_0}\right)\,\int\ud\Omega_2
\,\varphi_3(\mathbf{n})\,,
\end{equation}
where $\ud\Omega_2$ denotes the standard surface element in the
direction $\mathbf{n}$. As we see the crucial coefficient in the
expansion~\eqref{Fdev} is that for $p=3$; it corresponds to a
logarithmic divergence of the original integral~\eqref{IcalR}.

In the present paper, motivated by the success of dimensional regularization
when treating the UV divergencies, we treat the IR divergences of the
integral~\eqref{I} with the same regularization procedure. In $d$ spatial
dimensions the equivalent of $F(\mathbf{x})$, \textit{i.e.}, arising from the
same PN iteration of the field equations but performed in $d$ dimensions, will
be a function $F^{(d)}(\mathbf{x})$ with a more general expansion when
$r\to +\infty$ of the type\footnote{In Appendix~\ref{app:mult} we shall refer
  to the far zone expansion when $r\to +\infty$ as a ``multipole'' expansion
  and conveniently denote it as $\mathcal{M}(F^{(d)})$.}
\begin{equation}\label{Fdevddim}
F^{(d)}(\mathbf{x}) = \sum_{p\geqslant
  -p_0}\sum_{q=-q_0}^{q_1}\frac{1}{r^{p}}\left(\frac{\ell_0}{r}\right)^{q
  \varepsilon}\varphi^{(\varepsilon)}_{p,q}(\mathbf{n})\,.
\end{equation}
The difference with~\eqref{Fdev} is that the powers of $1/r$ now
depend linearly on $\varepsilon=d-3$, with $p\in\mathbb{Z}$ as before
and with also $q\in\mathbb{Z}$, bounded from below and from above
by $-q_0$ and $q_1$. Here $\ell_0$ denotes the usual constant scale
associated with dimensional regularization. Assuming that the
coefficients $\varphi^{(\varepsilon)}_{p,q}$ have a well-defined limit
when $\varepsilon\to 0$, \textit{i.e.}, that they do not contain any
pole $\propto 1/\varepsilon$ (such an assumption is always verified at
4PN order), we obtain the following relation with the coefficients
$\varphi_{p}$ in the limit $\varepsilon\to 0$,
\begin{equation}\label{relcoeff}
\varphi_{p}(\mathbf{n}) = \sum_{q=-q_0}^{q_1}
\varphi^{(\varepsilon=0)}_{p,q}(\mathbf{n})\,.
\end{equation}

The dimensional regularization (DR) prescription, to be considered as
usual in the sense of complex analytic continuation in
$d\in\mathbb{C}$, reads now
\begin{equation}\label{IDRdef}
I_\mathcal{R}^\text{DR} = \int_{r>\mathcal{R}}
\frac{\ud^{d}\mathbf{x}}{\ell_0^{d-3}}\,F^{(d)}(\mathbf{x}) \,.
\end{equation}
Working in the limit where $\varepsilon\to 0$, \textit{i.e.}, keeping
only the pole $\propto 1/\varepsilon$ followed by the finite part
$\propto \varepsilon^0$, and using also the relation~\eqref{relcoeff},
we readily obtain\footnote{\textit{A priori} the result also contains
  terms that diverge at infinity. These terms correspond to the
  coefficients $\varphi^{(\varepsilon)}_{p,q}$ with $q=1$ and
  $p\leqslant 3$, but do not appear in our computation.}
\begin{equation}\label{IDR}
I_\mathcal{R}^\text{DR} = - \sum_{p\not=
  3}\frac{\mathcal{R}^{3-p}}{3-p}\,\int\ud\Omega_2\,\varphi_p(\mathbf{n})
+ \sum_q\left[\frac{1}{(q-1)\varepsilon} -
  \ln\left(\frac{\mathcal{R}}{\ell_0}\right)\right]
\int\ud\Omega_{2+\varepsilon}\,\varphi^{(\varepsilon)}_{3,q}(\mathbf{n})
+ \mathcal{O}\left(\varepsilon\right)\,.
\end{equation}
Very important in this formula, is that the angular integration in the
second term, because of the presence of the pole, is to be performed
over the $(d-1)$-dimensional sphere, with surface element
$\ud\Omega_{2+\varepsilon}(\mathbf{n})$, up to order $\varepsilon$.

We shall thus add to the computations of Papers~I and~II the
\textit{difference} between the two prescriptions, say $\mathcal{D}I =
I_\mathcal{R}^\text{DR}-I_\mathcal{R}^\text{HR}$. Note that the first
term in~\eqref{IDR} is identical to the corresponding term
in~\eqref{IHR}, and thus cancels out in the difference. We thus
obtain, to dominant order when $\varepsilon\to 0$,
\begin{equation}\label{resdiff}
\mathcal{D}I = \sum_q\left[\frac{1}{(q-1)\varepsilon} -
  \ln\left(\frac{r_0}{\ell_0}\right)\right]
\int\ud\Omega_{2+\varepsilon}\,\varphi^{(\varepsilon)}_{3,q}(\mathbf{n})
+ \mathcal{O}\left(\varepsilon\right)\,,
\end{equation}
where, as expected, the scale $\mathcal{R}$ has disappeared from the
difference. 

We have applied the formula~\eqref{resdiff} to each of the terms
composing the gravitational part $L_g$ of the Fokker Lagrangian. Thus,
we have computed the expansion when $r\to +\infty$ of the various
potentials parametrizing the metric in $d$ dimensions as given by
Eqs.~(4.14)--(4.15) in Paper~I.\footnote{Extensive use is made of the
  software Mathematica together with the tensor package
  \textit{xAct}~\cite{xtensor}.} These potentials are those needed at
the 4PN order following the method ``$n+2$'' described in Sec.~IVA of
Paper~I. For this calculation we use the far-zone expansion of some
elementary functions in $d$ dimensions (notably the elementary Fock
kernel $g$~\cite{Fock}); this will be described in
Appendix~\ref{app:mult}. Once we have computed the expansions of all
the potentials we plug them into $L_g$ and obtain the coefficients
$\varphi^{(\varepsilon)}_{3,q}(\mathbf{n})$ corresponding to all the
terms. Then we simply evaluate Eq.~\eqref{resdiff} for each of the
terms\footnote{In practical calculations we always verify that the
  coefficient $\varphi^{(\varepsilon)}_{3,1}(\mathbf{n})$ averages to
  zero, so that there is no problem with the value $q=1$ in
  Eq.~\eqref{resdiff}.} and obtain the Fokker action with IR
divergences correctly regularized by means of DR.

The total difference will actually be called
$\mathcal{D}L^\text{inst}_g=\sum\mathcal{D}I$. Indeed it is composed
of all the terms obtained following the method $n+2$, which keeps
track of the ``instantaneous'' terms, but neglects the ``tail'' term
which will be investigated in Sec.~\ref{sec:tail}. Thus,
$\mathcal{D}L^\text{inst}_g$ is composed of a pole part $\propto
1/\varepsilon$ followed by a finite part $\propto \varepsilon^0$ which
depends on the arbitrary IR scale $r_0$ as well as on $\ell_0$. We
next look for a (physically irrelevant) shift that will remove most of
the poles $1/\varepsilon$ and eliminate most of the dependence on the
constant $r_0$. We find, after applying a suitable shift, that the
difference becomes (irreducibly)
\begin{align}\label{DLginst}
\mathcal{D}L^\text{inst}_g &= \frac{G^2 m}{5 c^8}
\left[\frac{1}{\varepsilon} -2
  \ln\left(\frac{\sqrt{\bar{q}}\,r_0}{\ell_0}\right)\right]
\left(I^{(3)}_{ij}\right)^2 \nonumber\\ &+
\frac{G^4m\,m_1^2m_2^2}{c^8r_{12}^4}\biggl( -
\frac{2479}{150}(n_{12}v_{12})^2 + \frac{1234}{75} v_{12}^2 \biggr) +
\mathcal{O}\left(\varepsilon\right)\,,
\end{align}
where we pose $\bar{q}=4\pi\ue^{\gamma_\text{E}}$ with $\gamma_\text{E}$
being the Euler constant. The other notations are exactly the same as
in Papers~I and~II, \textit{e.g.}, $m=m_1+m_2$ is the total mass and
$(n_{12}v_{12})$ is the Euclidean scalar between the relative
direction between the two bodies and their relative velocity.

As we see there is a remaining pole in Eq.~\eqref{DLginst}, and we
shall prove in Sec.~\ref{sec:tail} that it will be cancelled by a
corresponding pole coming from the 4PN tail term evaluated in $d$
dimensions. The pole is proportional to the square of the third
time-derivative of the quadrupole moment $I_{ij}$. In a small 4PN
term, the quadrupole can be taken to be the Newtonian one; however,
consistently with the pole $1/\varepsilon$ in front, it is to be
evaluated in $d$ dimensions, up to order $\varepsilon$ included. For
completeness we show here the complete expression up to that order,
\begin{align}\label{Iij3carre}
\left(I^{(3)}_{ij}\right)^2 &= \frac{G^3 m_1^2m_2^2}{r_{12}^4}\biggl(
- \frac{88}{3}(n_{12}v_{12})^2 + 32 v_{12}^2 \biggr) \left[ 1 -
  \frac{\varepsilon}{2}
  \ln\left(\frac{\sqrt{\bar{q}}\,r_{12}}{\ell_0}\right)\right]
\nonumber\\ &+ \varepsilon \frac{G^3 m_1^2m_2^2}{r_{12}^4}\biggl( -
\frac{836}{9}(n_{12}v_{12})^2 + 96 v_{12}^2 \biggr) +
\mathcal{O}\left(\varepsilon^2\right)\,.
\end{align}

Gladly, we discover that the two terms in the second line of
Eq.~\eqref{DLginst} have exactly the structure of the two
``ambiguity'' parameters $\delta_1$ and $\delta_2$ that were
introduced in Paper~II. As we shall see, this will permit to confirm
the conjecture advocated in Paper~II, namely that different IR
regularizations have merely the effect of changing the values of two
and only two ambiguity parameters $\delta_1$ and $\delta_2$ (modulo,
of course, irrelevant world-line shifts).

Next, in addition to Eq.~\eqref{DLginst}, we must also consider
another ``instantaneous'' contribution when working in full DR. This
is due to the fact that in HR it matters if we start from a
gravitational Lagrangian at quadratic order of the type $\sim\partial
h\partial h$ or of the type $\sim h \Box h$ (\textit{i.e.}, the
propagator form). Indeed, the two Lagrangians differ by a surface term
$\sim\partial(h\partial h)$ coming from the integration by part, and
this surface term does contribute in HR. On the contrary, in DR it
does not matter whether one starts with the Lagrangian in the form
$\sim\partial h\partial h$ or with the Lagrangian in propagator form
$\sim h\Box h$ because the surface term is always zero in DR by
analytic continuation in the dimension $d$. The fact that the two
Lagrangians are equivalent in DR constitutes a very nice feature of DR
as opposed to HR. In Paper~I we initially performed our HR calculation
with the $\sim\partial h\partial h$ Lagrangian and then corrected it
by adding the appropriate surface term so that our HR prescription
starts with a Lagrangian having the propagator form $\sim h \Box
h$. On the other hand our calculation of the difference
yielding~\eqref{DLginst} has been done with the prescription
$\sim\partial h\partial h$, so we now have to subtract off the latter
surface term. After applying an appropriate shift, this gives the
following contribution to be \textit{subtracted} from the HR result in
order to control the full DR:
\begin{equation}\label{DLgsurf}
\mathcal{D}L^\text{surf}_g =
\frac{G^4m\,m_1^2m_2^2}{c^8r_{12}^4}\biggl[ -
  \frac{52}{15}(n_{12}v_{12})^2 + \frac{64}{15} v_{12}^2 \biggr] \,.
\end{equation}
Again we find it to have the form of the ambiguity parameters modulo
shifts.

In the language of EFT (see for instance Ref.~\cite{PR17}) our
``instantaneous'' calculation which has been done in the present
section and yields Eq.~\eqref{DLginst}, corresponds to the so-called
``potential mode'' contribution, say $V_\text{pot}$. As emphasized
in~\cite{PR17,Po17}, the pole it contains is an IR pole, thus
$\varepsilon\equiv\varepsilon_\text{IR}$. However, there is now to
take into account the contribution coming from the conservative part
of the 4PN tail effect in $d$ dimensions, which corresponds in the EFT
language to the ``radiation'' contribution, say $V_\text{rad}$. As we
shall prove in Sec.~\ref{sec:tail} the IR pole in Eq.~\eqref{DLginst}
will be cancelled by a corresponding UV pole
$\varepsilon\equiv\varepsilon_\text{UV}$ coming from the radiation
term in $d$ dimensions.

\section{Formulas for the near-zone expansion in $d$ dimensions} 
\label{sec:NZ}

In this section and the following one we shall prove that there is
another contribution in the difference between DR and HR, coming from
the tail effect in $d$ dimensions. Indeed the computation in the
previous section was based on the method ``$n+2$'' (see Sec.~IV A of
Paper~I) which is valid for \textit{symmetric} terms defined from the
usual symmetric propagator. However the tail effect at 4PN order is to
be added separately since it is in the form of an hereditary type
homogeneous solution of the wave equation, which is of the
\textit{anti-symmetric} type (\textit{i.e.}, advanced minus retarded),
thus regular when $r\to 0$, and which has not been taken into account
in the method $n+2$.

We start by general considerations on the near-zone expansion of the
solution of the flat scalar wave equation in $d+1$ space-time
dimensions (thus, with $\mathbf{x}\in\mathbb{R}^d$),\footnote{General
  conventions from earlier works~\cite{BDE04,BDEI05dr} are adopted. We
  pose $G=c=1$ in this section.}
\begin{equation}\label{waveeq}
\Box h(\mathbf{x},t) = N(\mathbf{x},t)\,.
\end{equation}
The source of such an equation will represent a generic term in the source of
the equation~\eqref{quadratic} that we shall solve in the next section. The
retarded Green's function $G_\text{ret}(\mathbf{x},t)$ of that scalar wave
equation, thus satisfying
$\Box G_\text{ret}(\mathbf{x},t) = \delta(t)\delta^{(d)}(\mathbf{x})$,
explicitly reads
\begin{equation}\label{Gret}
G_\text{ret}(\mathbf{x},t) = -
\frac{\tilde{k}}{4\pi}\frac{\theta(t-r)}{r^{d-1}}
\,\gamma_{\frac{1-d}{2}}\left(\frac{t}{r}\right)\,,
\end{equation}
where $\tilde{k}=\pi^{1-\frac{d}{2}}\Gamma(\frac{d}{2}-1)$ (with
$\Gamma$ being the usual Eulerian function) denotes a pure constant so
defined that $\lim_{d\to3}\tilde{k}=1$, and $\theta(t-r)$ denotes the
usual Heaviside step function. The corresponding advanced Green's
function $G_\text{adv}(\mathbf{x},t)$ is given by the same expression
but with $\theta(-t-r)$ instead of $\theta(t-r)$. We have introduced
for convenience the function $\gamma_s(z)$ defined for any
$s\in\mathbb{C}$ and $\vert z\vert\geqslant 1$ by\footnote{The
  function $\gamma_s(z)$ is the natural generalization of the function
  $\gamma_\ell(z)$ (for $\ell\in\mathbb{N}$) introduced
  in~\cite{PB02,BFN05} to parametrize ``radiation-reaction'' STF
  multipole moments. In a similar way one can introduce a function
  $\delta_s(z)$ which would be a generalization of the function
  parametrizing the ``source-type'' multipole moments~\cite{B98mult},
\begin{equation*}
\delta_s(z) = \frac{\Gamma(s+\frac{3}{2})}{\sqrt{\pi}\Gamma(s+1)}
\,\big(1-z^2\bigr)^s\,,
\end{equation*}
and satisfying $\int_{-1}^{1} \ud z \,\delta_s(z) = 1$. One can show
that $\gamma_s(z)=-(1+\ue^{-2\ui \pi s})\delta_s(z)$, thus
$\gamma_\ell(z)=-2\delta_\ell(z)$ when $\ell\in\mathbb{N}$. Note also
that the Riesz~\cite{Riesz} kernels $Z_\alpha(t,r)$ in Minkowski $d+1$
space-time (satisfying the convolution algebra $Z_\alpha * Z_\beta =
Z_{\alpha+\beta}$) are given in terms of the function $\gamma_s(z)$ by
\begin{equation*}
Z_\alpha(t,r) =
\frac{\Gamma(\frac{d-\alpha}{2})}{\Gamma(\frac{\alpha}{2})}
\,\frac{r^{\alpha-d-1}}{2^\alpha
  \pi^{\frac{d}{2}}}\,\gamma_{\frac{\alpha-d-1}{2}}\!\left(\frac{t}{r}\right)\,.
\end{equation*}}
\begin{align}\label{gammas}
\gamma_s(z) &= \frac{2\sqrt{\pi}}{\Gamma(s+1)\Gamma(-s-\frac{1}{2})}
\,\big(z^2-1\bigr)^s \nonumber\\ &=
\frac{\Gamma(-s)}{2^{2s+1}\Gamma(s+1)\Gamma(-2s-1)}
\,\big(z^2-1\bigr)^s\,,
\end{align}
where the normalisation has been chosen so that
\begin{equation}\label{norm}
\int_1^{+\infty} \ud z \,\gamma_s(z) = 1\,.
\end{equation}
The latter integral converges when $-1 < \Re(s) < - \frac{1}{2}$ and
can be extended to any $s\in\mathbb{C}$ by complex analytic
continuation. For strictly negative integer values (say $s\in
-1-\mathbb{N}$) the function~\eqref{gammas} is zero in an ordinary
sense, but is actually a distribution; for instance we can check that
$\gamma_{-1}(z)=\delta(z-1)$ (see Appendix~\ref{app:distr}). Notice
that the Green's function~\eqref{Gret} depends only on $t$ and the
$d$-dimensional Euclidean norm $r=\vert\mathbf{x}\vert$. Its Fourier
transform is also known [see \textit{e.g.} Eq.~(2.4) in
Ref.~\cite{BDEI05dr}]. The retarded solution of the wave
equation~\eqref{waveeq} is given by
\begin{align}\label{retsol}
h(\mathbf{x},t) &= \int_{-\infty}^{+\infty}\ud t'\int
\ud^d\mathbf{x}'\,G_\text{ret}(\mathbf{x}-\mathbf{x}',t-t')
\,N(\mathbf{x}',t') \nonumber\\ &=- \frac{\tilde{k}}{4\pi}
\int_1^{+\infty} \ud z \,\gamma_{\frac{1-d}{2}}(z) \int
\ud^d\mathbf{x}'
\,\frac{N(\mathbf{x}',t-z\vert\mathbf{x}-\mathbf{x}'\vert)}
       {\vert\mathbf{x}-\mathbf{x}'\vert^{d-2}}\,.
\end{align}

Now, we want to identify a piece in this solution, that will be a
homogeneous anti-symmetric solution of the wave equation which is
regular when $\mathbf{x} \to 0$. It may be obtained by performing the
formal near-zone expansion of $h(\mathbf{x},t)$. Later we shall
use this homogeneous solution to control the tail effect in the near
zone. Thus, for this application we consider that $N(\mathbf{x},t)$
represents a particular term in the quadratic part of the Einstein
field equations outside the matter source, \textit{i.e.}, a generic
term of $N_2[h_1]$ in Eq.~\eqref{quadratic} below. In particular
$N(\mathbf{x},t)$ is to be thought as already ``multipole-expanded''
outside the matter source.

We start from Eq.~\eqref{retsol} in which we swap the time and space
integrals, defining
\begin{equation}\label{Stildedef}
\tilde{N}_\text{ret}(\mathbf{x}',\vert\mathbf{x}-\mathbf{x}'\vert,t)
=- \frac{\tilde{k}}{4\pi} \int_1^{+\infty} \ud z
\,\gamma_{\frac{1-d}{2}}(z)
\,\frac{N(\mathbf{x}',t-z\vert\mathbf{x}-\mathbf{x}'\vert)}
       {\vert\mathbf{x}-\mathbf{x}'\vert^{d-2}}\,,
\end{equation}
which is a homogeneous solution of the wave equation with respect to the field
point $\mathbf{x}$:
$\Box\tilde{N}_\text{ret}(\mathbf{x}',\vert\mathbf{x}-\mathbf{x}'\vert,t) =
0$.
Homogeneous solutions of the wave equation are investigated in general terms
in Appendix~\ref{app:hom}. The $d$-dimensional integral~\eqref{retsol} is
defined by complex analytic continuation in $d=3+\varepsilon$, and we are
looking to the neighbourhood of $\varepsilon=0$, the latter point being
excluded. However, we shall find that for some particular terms in our
calculation, the analytic continuation cannot be performed as the
$\varepsilon$'s cancel out. In order to be protected when such a cancellation
happens, we introduce a regulator ${r'}^\eta$ in factor of the source (where
$r'=\vert\mathbf{x}'\vert$), and carry out all calculations with some finite
parameter $\eta\in\mathbb{C}$, invoking the analytic continuation in $\eta$
when necessary. At the end of our calculation we shall compute the limit when
$\eta\to 0$, and find that this limit is finite for any $\varepsilon$. Finally
we apply the DR prescription on the result, looking for the neighbourhood of
$\varepsilon=0$ and the presence of poles $1/\varepsilon$. For more details see the companion paper~\cite{MBBF17}, where this method is referred to as the ``$\varepsilon \eta$'' regularization. From now on we thus
consider (with implicit limit when $\eta\to 0$)
\begin{equation} \label{phiFP}
h(\mathbf{x},t) = \int \ud^d\mathbf{x}'\,{r'}^\eta\,\tilde{N}_\text{ret}
  (\mathbf{x}',\vert\mathbf{x}-\mathbf{x}'\vert,t) \,. \end{equation}

The $d$-dimensional integral is then split into two pieces, each of
which corresponds to the regions of integration $|\mathbf{x}'| < R$
and $\vert\mathbf{x}'\vert > R$, respectively, for some positive
$R$. If we choose $R$ equal to the near-zone radius, we are allowed to
replace the source
$\tilde{N}_\text{ret}(\mathbf{x}',\vert\mathbf{x}-\mathbf{x}'\vert,t)$
of the inner integral by its own PN expansion, as given by
Eqs.~\eqref{even}--\eqref{odd} in Appendix~\ref{app:hom}. The result
may be written as an integral over the whole space, minus the same
integral over the region $\vert\mathbf{x}'\vert < R$. This yields
\begin{align}
h(\mathbf{x},t) &= \int \ud^d\mathbf{x}'
\,{r'}^\eta\,\overline{\tilde{N}_\text{ret}
  (\mathbf{x}',\vert\mathbf{x}-\mathbf{x}'\vert,t)} \nonumber \\ & +
\int_{\vert \mathbf{x}'\vert>R} \ud^d\mathbf{x}'
\,{r'}^\eta\,\left[\tilde{N}_\text{ret}
  (\mathbf{x}',\vert\mathbf{x}-\mathbf{x}'\vert,t) -
  \overline{\tilde{N}_\text{ret}
    (\mathbf{x}',\vert\mathbf{x}-\mathbf{x}'\vert,t)}\right]\,, 
\label{decompNZFZ}
\end{align}
where the overbar refers to the PN expansion. Next, in the second
integral, extending over the exterior zone
($\vert\mathbf{x}'\vert>R$), we can perform a formal Taylor expansion
when $\vert\mathbf{x}'\vert \to +\infty$. After expressing the result
in terms of symmetric-trace-free (STF) tensors, we find
\begin{equation}\label{taylor}
\tilde{N}_\text{ret} (\mathbf{x}',\vert\mathbf{x}-\mathbf{x}'\vert,t)
= \sum_{q=0}^{+\infty} \frac{(-)^q}{q!}  \sum_{j=0}^{+\infty}
\Delta^{-j} \hat{x}_Q \left(\hat{\partial}'_Q
\tilde{N}_\text{ret}^{(2j)}(\mathbf{y},r',t)\right)_{\mathbf{y}=\mathbf{x}'}\,,
\end{equation}
where $\hat{\partial}'_Q$ denotes the STF projection of a product of
$q$ partial derivatives
$\partial'_Q=\partial'_{i_1}\cdots\partial'_{i_q}$ with respect to
${x'}^i$ (\textit{i.e.}, $\partial'_i=\partial/\partial x'^i$), where
$Q=i_1\cdots i_q$ is a multi-index with $q$ indices, and where the
time multi-derivatives are indicated with the superscript index
$(2j)$. Furthermore we employ the useful short-hand notation (with
$r=\vert\mathbf{x}\vert$)~\cite{PB02,BFN05}
\begin{equation}\label{Deltaj}
\Delta^{-j} \hat{x}_Q =
\frac{\Gamma(q+\frac{d}{2})}{\Gamma(q+j+\frac{d}{2})} \,\frac{r^{2j}
  \hat{x}_Q}{2^{2j} j!}\,,
\end{equation}
for the iterated inverse Poisson operator acting on the STF product
$\hat{x}_Q$ of $q$ source points $x^i$, such a notation being
motivated by the fact that
$\Delta(\Delta^{-j} \hat{x}_Q)=\Delta^{-j+1} \hat{x}_Q$. Notice that
in Eq.~\eqref{taylor} the point $\mathbf{y}$ is held constant when
applying the partial derivatives, and is to be replaced by
$\mathbf{x}'$ only afterwards. The same treatment applies also for the
overbared quantity in the last term of~\eqref{decompNZFZ}. At this
stage we obtain the near-zone or PN expansion
\begin{align}
\overline{h(\mathbf{x},t)} &= \int \ud^d\mathbf{x}'
\,{r'}^\eta\,\overline{\tilde{N}_\text{ret}
  (\mathbf{x}',\vert\mathbf{x}-\mathbf{x}'\vert,t)} \nonumber \\ & +
\sum_{q=0}^{+\infty} \frac{(-)^q}{q!}  \sum_{j=0}^{+\infty}
\Delta^{-j} \hat{x}_Q \int_{\vert \mathbf{x}'\vert>R} \ud^d\mathbf{x}'
\,{r'}^\eta\,\left(\hat{\partial}'_Q \tilde{N}_\text{ret}^{(2j)} -
\overline{\hat{\partial}'_Q
  \tilde{N}_\text{ret}^{(2j)}}\right)_{\mathbf{y}=\mathbf{x}'}\,.
\label{atthisstage}
\end{align}
Applying the same idea as before, \textit{i.e.}, decomposing the
second term as an integral over the whole space minus the same
integral restricted to the inner region $\vert x'\vert<R$, we can
further rewrite the above expression as
\begin{equation}\label{decompNZFZ2}
\overline{h} = \int \ud^d\mathbf{x}'
\,{r'}^\eta\,\overline{\tilde{N}_\text{ret}
  (\mathbf{x}',\vert\mathbf{x}-\mathbf{x}'\vert,t)} +
\sum_{q=0}^{+\infty} \frac{(-)^q}{q!}  \sum_{j=0}^{+\infty}
\Delta^{-j} \hat{x}_Q \int \ud^d\mathbf{x}' \,{r'}^\eta\,\left(\hat{\partial}'_Q
\tilde{N}_\text{ret}^{(2j)}\right)_{\mathbf{y}=\mathbf{x}'} \!\! +
\overline{\Delta} \,.
\end{equation}
This takes almost the requested form, but there is still the last term
with a peculiar unwanted form, given by
\begin{equation}\label{termeenplus}
\overline{\Delta} = - \sum_{q=0}^{+\infty} \frac{(-)^q}{q!}
\sum_{j=0}^{+\infty} \Delta^{-j} \hat{x}_Q \left[\int_{\vert
    \mathbf{x}'\vert<R} \ud^d\mathbf{x}' \,{r'}^\eta\,\left(\hat{\partial}'_Q
  \tilde{N}_\text{ret}^{(2j)}\right)_{\mathbf{y}=\mathbf{x}'} +
  \int_{\vert \mathbf{x}'\vert>R} \ud^d\mathbf{x}'
  \,{r'}^\eta\,\left(\overline{\hat{\partial}'_Q
    \tilde{N}_\text{ret}^{(2j)}}\right)_{\mathbf{y}=\mathbf{x}'}\right]\,.
\end{equation}
However, in the near-zone integral, we can again replace the integrand
by the PN expansion, so that the two integrals combine to a single
integral extending over all space, which \textit{in fine} turns out to
be formally zero:
\begin{equation}\label{termeenplus0}
\overline{\Delta} = - \sum_{q=0}^{+\infty} \frac{(-)^q}{q!}
\sum_{j=0}^{+\infty} \Delta^{-j} \hat{x}_Q \int \ud^d\mathbf{x}'
\,{r'}^\eta\,\left(\overline{\hat{\partial}'_Q
  \tilde{N}_\text{ret}^{(2j)}}\right)_{\mathbf{y}=\mathbf{x}'} = 0\,.
\end{equation}
To prove the last statement we recall from
Eqs.~\eqref{even}--\eqref{odd} that
$\overline{\tilde{N}_\text{ret}(\mathbf{y},
  \vert\mathbf{x}'\vert,t)}$,
which is the PN expansion of a retarded solution of the wave equation,
has the form of a sum $\sum F_{a,b}(t) \,{r'}^{a+\varepsilon b}$.
Hence, when integrating this term and after performing the angular
integration, we find a radial integral of the type
$\int_0^{+\infty} \ud r'\,{r'}^{a'+b'\varepsilon+\eta}$, which is thus
zero by analytic continuation in $\varepsilon$, except for the
particular case where $b'=0$; the latter case is precisely the one
where we need the ``protection'' of the regulator ${r'}^\eta$ in order
to complete our proof. Not only is the regulator important for
establishing~\eqref{termeenplus0} but it permits a complete
calculation of all the terms (see the Appendix~\ref{app:coeffC}). We
shall find that the limit $\eta\to 0$ is perfectly well defined for
the sum of all the terms as the poles $1/\eta$ cancel out.

The first term on the right-hand side of~\eqref{decompNZFZ2}
represents the retarded integral acting directly on the PN expansion
of the source, \textit{i.e.}, $\overline{N}$ (or, rather,
${r}^\eta\overline{N}$). Thus the PN expansion of the corresponding
solution can now be rewritten as
\begin{equation}\label{PNsolexp}
  \overline{h} = \overline{\Box_\text{ret}^{-1}} \bigl[{r}^\eta
  \overline{N}\bigr] +
  \sum_{q=0}^{+\infty} \frac{(-)^q}{q!}  \sum_{j=0}^{+\infty}
  \Delta^{-j} \hat{x}_Q \int \ud^d\mathbf{x}' \,{r'}^\eta\,
   \left(\hat{\partial}'_Q
    \tilde{N}_\text{ret}^{(2j)}\right)_{\mathbf{y}=\mathbf{x}'} \,,
\end{equation}
where the retardations in the inverse d'Alembertian operator are PN expanded.
Since the first term is obviously a particular solution of the
(PN-expanded) wave equation in the limit $\eta\to 0$, the second term
in~\eqref{PNsolexp} is a homogeneous solution; let us call it
$\overline{h}^\text{asym}$ for a reason to soon become clear. In
more details it reads
\begin{equation}\label{varphiexp0}
\overline{h}^\text{asym} = - \frac{\tilde{k}}{4\pi}
\sum_{q=0}^{+\infty} \frac{(-)^q}{q!} \sum_{j=0}^{+\infty} \Delta^{-j}
\hat{x}_Q \int_1^{+\infty} \ud z \,\gamma_{\frac{1-d}{2}}(z) \,\int
\ud^d\mathbf{x}'\,{r'}^\eta\,\hat{\partial}'_Q\!\left[\frac{N^{(2j)}(\mathbf{y},t
    - z r')}{{r'}^{d-2}}\right]_{\mathbf{y}=\mathbf{x}'}\,.
\end{equation}
This is our looked-for homogeneous solution; it is clearly of the form
$\overline{h}^\text{asym} =
\sum_{q=0}^{+\infty}\sum_{j=0}^{+\infty} \Delta^{-j}
\hat{x}_Q\,F_Q^{(2j)}(t)$, on which form we can directly check that
$\Box \overline{h}^\text{asym} = 0$. Furthermore, that solution
is manifestly regular when $r\to 0$, and so it must be identified with
a homogeneous \textit{anti-symmetric} solution of the wave equation
in $d$ dimensions, of the type half-retarded minus advanced. In
particular, Eq.~\eqref{varphiexp0} must be identified with an
anti-symmetric solution $H^\text{asym}$ whose general form is given by
Eq.~\eqref{multanti}. Bearing unimportant factors, this means that we
should always be able to find a function $f_Q(t)$ such that
\begin{equation}\label{solvefF}
F_Q(t) = \int_0^{+\infty}\ud\tau\,\tau^{-\varepsilon}
\Bigl[f_Q^{(2\ell+2)}(t-\tau) - f_Q^{(2\ell+2)}(t+\tau)\Bigr]\,.
\end{equation}
We prove this statement by going to the Fourier domain. Given the
Fourier transform $\hat{F}_Q(\omega)$ of $F_Q(t)$, Eq.~\eqref{solvefF}
will be verified provided that the Fourier transform
$\hat{f}_Q(\omega)$ of $f_Q(t)$ takes the expression
\begin{equation}\label{Fourier}
\hat{f}_Q(\omega) = \frac{2\ui (-)^\ell}{\cos(\frac{\pi
    \varepsilon}{2})\Gamma(1-\varepsilon)} \frac{\text{sign}(\omega)}
    {\vert\omega\vert^{2\ell+1+\varepsilon}}\hat{F}_Q(\omega)\,.
\end{equation}

Next, we consider the case of a source term which has a definite
multipolarity $\ell$, namely $N(\mathbf{x},t) = \hat{n}_L N(r,t)$,
where $\hat{n}_L$ is the STF projection of the product of $\ell$ unit
vectors $n_i$, and like before $L=i_1\cdots i_\ell$. We shall denote
the corresponding solution by
$\overline{h}^\text{asym}_L(\mathbf{x},t)$. Using
$\hat{\partial}'_Qf(r') = \hat{n}'_Q {r'}^q({r'}^{-1}\ud/\ud r')^q
f(r')$ in~\eqref{varphiexp0}, we can explicitly perform the angular
integration in $d$ dimensions [see \textit{e.g.} Eqs.~(B23)
in~\cite{BDE04}], and get
\begin{align}\label{varphiexp1}
\overline{h}^\text{asym}_L &=
\frac{(-)^{\ell+1}\Gamma(\frac{d}{2})}{2^\ell (d-2)
  \Gamma(\frac{d}{2}+\ell)} \,\sum_{j=0}^{+\infty} \Delta^{-j}
\hat{x}_L \int_1^{+\infty} \ud z \,\gamma_{\frac{1-d}{2}}(z)
\nonumber\\ &\qquad \times \int_0^{+\infty} \ud
r'\,{r'}^{d+\ell-1+\eta}\left(\frac{1}{r'}\frac{\ud}{\ud r'}\right)^\ell
\!\left[\frac{N^{(2j)}(\vert\mathbf{y}\vert,t - z
    r')}{{r'}^{d-2}}\right]_{\vert\mathbf{y}\vert=r'}\,.
\end{align}
Still this formula can be substantially simplified by means of a
series of integrations by parts over the $z$-variable, and we nicely
obtain
\begin{equation}\label{varphiexp2}
\overline{h}^\text{asym}_L = - \frac{1}{d+2\ell-2}
\,\sum_{j=0}^{+\infty} \Delta^{-j} \hat{x}_L \,\int_1^{+\infty} \ud z
\,\gamma_{\frac{1-d}{2}-\ell}(z) \,\int_0^{+\infty} \ud
r'\,{r'}^{-\ell+1+\eta} \,N^{(2j)}(r',t-z r')\,.
\end{equation}

We now specialize Eq.~\eqref{varphiexp2} to the case of a source term
made of a quadratic interaction between a monopolar static solution
$\propto r^{d-2}$ and some homogeneous multipolar \textit{retarded}
solution, namely, a spatial multi-derivative of a monopolar retarded
solution [see Eq.~\eqref{Itilde}]. Indeed, such source term will be the one
we meet when computing the tail effect as seen in the near zone ($r\to
0$). Its generic form is of the type (with $\varepsilon=d-3$)
\begin{equation}\label{Stail}
N(r,t) = r^{-k-2\varepsilon} \,\int_1^{+\infty} \ud y
\,y^p\,\gamma_{-1-\frac{\varepsilon}{2}}(y)\,F(t-y r)\,,
\end{equation}
where $k, p \in\mathbb{N}$ and the function $F(t)$ stands for some
time derivative of a component of a multipole moment, namely the
source quadrupole moment $I_{ij}(t)$ that we shall consider in
Sec.~\ref{sec:tail}. Plugging~\eqref{Stail} into~\eqref{varphiexp2},
and performing the change of integration variable
$r'\rightarrow\tau=(y+z)r'$, we readily obtain
\begin{equation}\label{varphiexp3}
\overline{h}^\text{asym}_L = -
\frac{C_\ell^{p,k}}{2\ell+1+\varepsilon} \,\sum_{j=0}^{+\infty}
\Delta^{-j} \hat{x}_L \,\int_0^{+\infty} \ud
\tau\,\tau^{-\ell-k+1-2\varepsilon+\eta}\,F^{(2j)}(t-\tau)\,,
\end{equation}
with the following purely numerical coefficient (also depending on the
dimension)
\begin{equation}\label{coeffC}
C_\ell^{p,k} = \int_1^{+\infty} \!\ud y
\,y^p\,\gamma_{-1-\frac{\varepsilon}{2}}(y)\int_1^{+\infty} \!\ud z
\,(y+z)^{\ell+k-2+2\varepsilon-\eta}\,\gamma_{-\ell-1-\frac{\varepsilon}{2}}(z)\,.
\end{equation}

We are ultimately interested in the limit $\varepsilon\to 0$, but it
is clear that the integral over $\tau$ in~\eqref{varphiexp3} becomes
ill-defined in this limit because of the bound $\tau=0$ of the
integral. On the other hand since $F(t)$ is a time derivative of a
multipole moment, we can assume that it is zero in a neighbourhood of
$t=-\infty$ so there is no problem with the bound $\tau=+\infty$ of
the integral. We thus make explicit the generic presence of a pole
$\propto 1/\varepsilon$ when $\varepsilon\to 0$ by operating the
integral $\ell+k-1$ times by parts. In contrast with the IR pole in
Sec~\ref{sec:diffIR}, such a pole will be an UV-type pole,
$\varepsilon\equiv\varepsilon_\text{UV}$. All surface terms vanish by
analytic continuation in $\varepsilon$ and because $F(t-\tau)$ is zero
when $\tau\to\infty$, so we arrive at
\begin{equation}\label{varphifinal}
\overline{h}^\text{asym}_L =
\frac{(-)^{\ell+k}\,C_\ell^{p,k}}{2\ell+1+\varepsilon}
\,\frac{\Gamma(2\varepsilon-\eta)}{\Gamma(\ell+k-1+2\varepsilon-\eta)}
\,\sum_{j=0}^{+\infty} \Delta^{-j} \hat{x}_L \,\int_0^{+\infty} \ud
\tau\,\tau^{-2\varepsilon+\eta}\,F^{(2j+\ell+k-1)}(t-\tau)\,.
\end{equation}

Note the retarded character of this solution, which comes directly
from the retarded character of the source term postulated in
Eq.~\eqref{Stail}. In our approach, we are iterating the Einstein
field equations by means of retarded potentials. Thus, at some given
non-linear order, for instance quadratic, we obtain a retarded source
term which represents the \textit{physical} solution, containing both
conservative and radiation-reaction dissipative effects. Only at this
stage do we identify an ``anti-symmetric'' piece which is a part of
the physical retarded solution generated by that source term, and
which will be associated with the tail effect in the near zone.

The equation~\eqref{varphifinal} is our final formula for this section,
with which we can directly control the looked-for limit when
$\varepsilon\to 0$. In generic cases a pole $\sim 1/\varepsilon$ will show up, 
while the finite part
beyond the pole will contain an ordinary tail integral with the usual
logarithmic kernel. The numerical coefficient $C_\ell^{p,k}$ defined
by Eq.~\eqref{coeffC} is \textit{a priori} not trivial to control, but
fortunately we have found a way to compute it analytically as
described in Appendix~\ref{app:coeffC}.

\section{Derivation of the tail term in $d$ dimensions} 
\label{sec:tail}

We shall compute the tail term in $d$ dimensions directly in the near
zone metric of general matter sources, then obtain its contribution in
the equations of motion of compact binaries and finally in the Fokker
action. The Einstein field equations in harmonic gauge in the vaccum
region outside an isolated source read
\begin{subequations}\label{EFE}
\begin{align}
\Box h^{\mu\nu} &= \Lambda^{\mu\nu}[h]\,,\\\partial_\nu h^{\mu\nu} &=
0\,,
\end{align}
\end{subequations}
where $\Box$ is the flat d'Alembertian operator,
$h^{\mu\nu}=\sqrt{-g}g^{\mu\nu}-\eta^{\mu\nu}$ is the ``gothic''
metric deviation from flat space-time, and $\Lambda^{\mu\nu}$ denotes
the non-linear gravitational source term, which is at least quadratic
in $h$ and its derivatives. As we shall see, to control the 4PN tail
effect we can limit ourselves to the quadratic non-linear order, say
$h^{\mu\nu} = G h^{\mu\nu}_{1} + G^2 h^{\mu\nu}_{2} +
\mathcal{O}(G^3)$. Denoting by $N^{\mu\nu}[h]$ the quadratic piece in
the non-linear source term $\Lambda^{\mu\nu}$ the equations to be
solved are thus
\begin{equation}\label{quadratic} 
\Box h^{\mu\nu}_{2} = N^{\mu\nu}[h_{1}] \,,
\end{equation}
together with $\partial_\nu h^{\mu\nu}_2=0$. At this stage we know
that the tail effect is an interaction between the constant mass of
the system $M$ and its time-varying mass-type STF quadrupole moment
$I_{kl}(t)$. Accordingly the linearized metric is composed of two
pieces, say $h^{\mu\nu}_1 = h^{\mu\nu}_M + h^{\mu\nu}_{I_{kl}}$. The
static one corresponding to the mass reads
\begin{equation}\label{monopole} h^{00}_{M} = -4\tilde{M}\,,\qquad
h^{0i}_{M} = 0\,,\qquad h^{ij}_{M} = 0\,, 
\end{equation}
while the dynamical one for the quadrupole moment in harmonic gauge is
given by
\begin{subequations}\label{quadrupole}
\begin{align}
h^{00}_{I_{kl}} &= -2 \partial_{ij}\tilde{I}_{ij}\,,\\ h^{0i}_{I_{kl}}
&= 2 \partial_{j}\tilde{I}^{(1)}_{ij}\,,\\ h^{ij}_{I_{kl}} &= -2
\tilde{I}^{(2)}_{ij}\,.
\end{align}
\end{subequations}
We are essentially following the notation of Eqs.~(3.44)
in~\cite{BDEI05dr}. In particular we denote a homogeneous retarded
solution of the d'Alembertian equation as
\begin{align}\label{Itilde}
\tilde{I}_{ij}(t,r) &= - 4\pi \int_{-\infty}^{+\infty} \ud
t'\,G_\text{ret}(\mathbf{x},t-t') \,I_{ij}(t')\nonumber\\ &=
\frac{\tilde{k}}{r^{d-2}}\int_1^{+\infty} \ud
z\,\gamma_{\frac{1-d}{2}}(z)\,I_{ij}(t-z r)\,.
\end{align}
See the retarded Green's function of the d'Alembertian equation in
Eq.~\eqref{Gret} above. For the static mass this reduces to a
homogeneous solution of the Laplace equation,
\begin{equation}\label{Mtilde}
\tilde{M}(r) = - 4\pi \,M \int_{-\infty}^{+\infty} \ud
t'\,G_\text{ret}(\mathbf{x},t-t') = \frac{\tilde{k} M}{r^{d-2}}\,.
\end{equation}

The quadratic source term $N^{\mu\nu}[h_1]$ built out of the
linearized metrics~\eqref{monopole}--\eqref{quadrupole} reads
\begin{subequations}\label{N2harm}
\begin{align}
N^{00}_{M\times I_{kl}} &= - {h}^{00}_{M}
\partial_{00}{h}^{00}_{I_{kl}} - {h}^{ij}_{I_{kl}}
\partial_{ij}{h}^{00}_{M} - \frac{3d-2}{2(d-1)}
\partial_{i}{h}^{00}_{M} \partial_{i}{h}^{00}_{I_{kl}} +
\partial_{i}{h}^{00}_{M} \partial_{0}{h}^{0i}_{I_{kl}}\,,\\
{N}^{0i}_{M\times I_{kl}} &= - {h}^{00}_{M}
\partial_{00}{h}^{0i}_{I_{kl}} + \frac{d}{2(d-1)}
\partial_{i}{h}^{00}_{M} \partial_{0}{h}^{00}_{I_{kl}} +
\partial_{j}{h}^{00}_{M} \partial_{0}{h}^{ij}_{I_{kl}}
\nonumber\\ &\qquad\qquad + \partial_{j}{h}^{00}_{M} \left(
\partial_{i}{h}^{0j}_{I_{kl}} - \partial_{j}{h}^{0i}_{I_{kl}}
\right)\,,\\ 
{N}^{ij}_{M\times I_{kl}} &= - {h}^{00}_{M}
\partial_{00}{h}^{ij}_{I_{kl}} + \frac{d-2}{d-1}
\partial_{(i}{h}^{00}_{M} \partial_{j)}{h}^{00}_{I_{kl}} -
\frac{d-2}{2(d-1)}\delta_{ij} \partial_{k}{h}^{00}_{M}
\partial_{k}{h}^{00}_{I_{kl}} \nonumber\\ &\qquad\qquad - \delta_{ij}
\partial_{k}{h}^{00}_{M} \partial_{0}{h}^{0k}_{I_{kl}} + 2
\partial_{(i}{h}^{00}_{M}\partial_{0}{h}^{j)0}_{I_{kl}}\,.
\end{align}
\end{subequations}
As we have investigated in Sec.~\ref{sec:NZ}, the tail effect we are
looking for comes from a suitable homogeneous anti-symmetric solution
of the wave equations~\eqref{quadratic}. We have therefore applied our
end result given by Eq.~\eqref{varphifinal}, together with the
explicit method for the computation of the coefficients $C_\ell^{p,k}$
as explained in Appendix~\ref{app:coeffC}, to each of the terms of
Eqs.~\eqref{N2harm}. We consider only the pole part $\propto
1/\varepsilon$ followed by the finite part when $\varepsilon\to 0$,
and re-expand when $c\to+\infty$ in order to keep only the terms
contributing at the 4PN order. We then obtain the homogeneous solution
responsible for the tails as\footnote{We suppress the mention
  ``$M\times I_{kl}$'', and restore the factors of $c$ and $G$. Here
  $G$ denotes the usual Newtonian constant, such that $G^{(d)} =
  G\,\ell_0^{d-3}$ in $d$ dimensions. We recall also that
  $\bar{q}=4\pi\ue^{\gamma_\text{E}}$.}
\begin{subequations}\label{h2harmhom}
\begin{align}
{h}^{00ii}_\text{asym} &= \frac{8 G^2M}{15 c^{10}} \,x^{ij}
\int_0^{+\infty}\ud\tau\left[\ln\left(\frac{c\sqrt{\bar{q}}\,\tau}{2\ell_0}
  \right) - \frac{1}{2\varepsilon} +
  \frac{61}{60}\right]I^{(7)}_{ij}(t-\tau) +
\mathcal{O}\left(\frac{1}{c^{12}}\right) \,,\\
{h}^{0i}_\text{asym} &= - \frac{8 G^2M}{3 c^{9}} \,x^j
\int_0^{+\infty}\ud\tau\left[\ln\left(\frac{c\sqrt{\bar{q}}\,\tau}{2\ell_0}
  \right) - \frac{1}{2\varepsilon} +
  \frac{107}{120}\right]I^{(6)}_{ij}(t-\tau) +
\mathcal{O}\left(\frac{1}{c^{11}}\right) \,,\\
{h}^{ij}_\text{asym} &= \frac{8 G^2M}{c^{8}}
\int_0^{+\infty}\ud\tau\left[\ln\left(\frac{c\sqrt{\bar{q}}\,\tau}{2\ell_0}
  \right) - \frac{1}{2\varepsilon} +
  \frac{4}{5}\right]I^{(5)}_{ij}(t-\tau) +
\mathcal{O}\left(\frac{1}{c^{10}}\right) \,,
\end{align}
\end{subequations}
where we have introduced the usual variable
${h}^{00ii} = \frac{2}{d-1}[(d-2){h}^{00} + {h}^{ii}]$ (see Paper~I). In this
standard harmonic gauge the tail integrals and their associated (UV) poles are
spread out in all components of the metric. In Eqs.~\eqref{h2harmhom} we have inserted the correct numerical coefficients computed in the companion paper~\cite{MBBF17}, which are crucial in the end in order to obtain the ``first'' ambiguity.

Alternatively, we can do the calculation starting from the linear
quadrupole metric in a transverse-tracefree (TT) harmonic gauge. Thus,
instead of Eqs~\eqref{quadrupole}, we may consider the linear
quadrupole TT metric
\begin{subequations}\label{quadrupoleTT}
\begin{align}
{h'}^{00}_{I_{kl}} &= 0\,,\\ {h'}^{0i}_{I_{kl}} &=
0\,,\\ {h'}^{ij}_{I_{kl}} &= -2 \tilde{I}^{(2)}_{ij}
+4\partial_{k(i}\tilde{I}_{j)k} - \frac{2}{d-1}\delta_{ij}
\partial_{kl}\tilde{I}_{kl} -
2\frac{d-2}{d-1}\partial_{ijkl}\tilde{I}^{(-2)}_{kl}\,. 
\label{quadrupoleTTij}
\end{align}
\end{subequations}
In the TT gauge the quadratic source term is especially simple,
\begin{subequations}\label{N2TT}
\begin{align}
{N'}^{00}_{M\times I_{kl}} &= - \partial_{ij}{h}^{00}_{M}
{h'}^{ij}_{I_{kl}}\,,\\
{N'}^{0i}_{M\times I_{kl}} &= \partial_{j}{h}^{00}_{M}
\partial_{0}{h'}^{ij}_{I_{kl}}\,,\\
{N'}^{ij}_{M\times I_{kl}} &= - {h}^{00}_{M}
\partial_{00}{h'}^{ij}_{I_{kl}}\,,
\end{align}
\end{subequations}
and, relaunching our calculation (with inputs from~\cite{MBBF17}), we readily obtain 
\begin{subequations}\label{h2TThom}
\begin{align}
{h'}^{00ii}_\text{asym} &= - \frac{2}{15} \frac{G^2M}{c^{10}}
\,x^{ij} I^{(6)}_{ij}(t) +
\mathcal{O}\left(\frac{1}{c^{12}}\right)\,,\\
{h'}^{0i}_\text{asym} &=
\frac{4}{5} \frac{G^2M}{c^{9}} \,x^{j} I^{(5)}_{ij}(t) +
\mathcal{O}\left(\frac{1}{c^{11}}\right)\,,\\
{h'}^{ij}_\text{asym} &=
\frac{16}{5} \frac{G^2M}{c^{8}}
\int_0^{+\infty}\ud\tau\left[\ln\left(\frac{c\sqrt{\bar{q}}\,\tau}{2\ell_0}
  \right) - \frac{1}{2\varepsilon} +
  \frac{9}{40}\right]I^{(5)}_{ij}(t-\tau) +
\mathcal{O}\left(\frac{1}{c^{10}}\right) \,.
\end{align}
\end{subequations}
In the TT gauge the tail integral and the associated pole appear only
in the spatial components of the metric (notice also that
${h'}^{ii}_\text{asym}=0$ in this case).

Finally the tails in the harmonic metric~\eqref{h2harmhom} or its TT
counterpart~\eqref{h2TThom} will yield a modification of the equations
of motion. To compute it in the simplest way we perform a gauge
transformation (this time, at quadratic order), so designed as to
transfer all relevant terms in the ``$00ii$'' component of the
metric. In the new gauge the 4PN tail effect is thus entirely
described by the single scalar potential ${h''}^{00ii}_\text{asym}$,
or equivalently by the $00$ component of the usual covariant metric,
given by ${g''}_{00}^\text{asym} =
-\frac{1}{2}{h''}^{00ii}_\text{asym}$. We finally obtain
\begin{equation}\label{g00}
{g''}_{00}^\text{asym} = - \frac{8G^2 M}{5c^{8}}\,x^{ij}
\int_0^{+\infty}\ud\tau\left[\ln\left(\frac{c\sqrt{\bar{q}}\,\tau}
  {2\ell_0}\right) - \frac{1}{2\varepsilon} + \frac{41}{60}
  \right]I^{(7)}_{ij}(t-\tau) +
\mathcal{O}\left(\frac{1}{c^{10}}\right) \,.
\end{equation}
This result properly recovers the known tail integral in 3 dimensions [see
Eqs.~(5.24) in~\cite{BD88}]. In addition, there appear the pole and a certain
numerical coefficient, say $\kappa=\frac{41}{60}$. The (UV-type) pole is in
agreement with the result of Ref.~\cite{GLPR16}. On the other hand the
constant $\kappa=\frac{41}{60}$ has the form of the ambiguity $\alpha$
introduced in Paper~I, which is itself equivalent to the ambiguity $C$
of the Hamiltonian formalism~\cite{DJS14}, and is now determined. Note that the value of
this constant is the same in both our calculations, in harmonic and TT gauge.

As shown here and in the companion paper~\cite{MBBF17}, the value of the numerical coefficient,
\textit{i.e.}, $\kappa=\frac{41}{60}$, comes out directly from the matching
equation between the near zone and the radiation zone, and a consistent application of the $\varepsilon\eta$ regularization together with the closed form expressions of the coefficients $C_\ell^{p,k}$ obtained in Appendix~\ref{app:coeffC}. Of course the value of $\kappa$ was already known from
comparison with GSF calculations, but we now directly obtain the correct value (see Sec.~\ref{sec:amb}), which also agrees with
the published coefficient obtained in the computation of the $d$-dimensional
tail effect by Galley \textit{et al}~\cite{GLPR16} using EFT methods [see
their Eq.~(3.3)].

Once we have the single scalar effect~\eqref{g00} at the level of the
metric, it is straightforward to obtain the equivalent effect at the
level of the Lagrangian or Fokker action. Recall that the
corresponding piece in the Fokker action will describe only the
conservative part of the dynamics associated with the tail effect (see
Paper~I for discussion). We thus find the manifestly time-symmetric
contribution to the gravitational part of the action,
\begin{equation}\label{StailDR1}
S_g^\text{tail} = \frac{G^2M}{5 c^8} \int_{-\infty}^{+\infty}\ud t
\,I_{ij}^{(3)}(t) \int_0^{+\infty} \ud\tau \left[
  \ln\left(\frac{c\sqrt{\bar{q}}\,\tau}{2\ell_0}\right) -
  \frac{1}{2\varepsilon} + \frac{41}{60} \right]\left(I_{ij}^{(4)}(t-\tau) -
  I_{ij}^{(4)}(t+\tau)\right) \,,
\end{equation}
which can elegantly be rewritten by means of the Hadamard partie finie
(Pf) integral as
\begin{equation}\label{StailDR2}
S_g^\text{tail} = \frac{G^2M}{5 c^8} \,\mathop{\text{Pf}}_{\tau_0^\text{DR}}
\int\!\!\!\int \frac{\ud t\ud t'}{\vert t-t'\vert}
I_{ij}^{(3)}(t)\,I_{ij}^{(3)}(t')\,,
\end{equation}
where $\tau_0^\text{DR} = \frac{2\ell_0}{c
  \sqrt{\bar{q}}}\,\ue^{\frac{1}{2\varepsilon}-\frac{41}{60}}$ plays the role
of the Hadamard cut-off scale. Finally, when considering the
difference between the DR and HR results, we have to correct for the
different treatments of the tail term in the two procedures. In
Sec.~III of Paper~I we obtained the tail term in the same form as
Eq.~\eqref{StailDR2} but with a different Hadamard scale
$\tau_0^\text{HR}=2s_0$. The difference of Lagrangians to be added to
the result of Paper~I concerning the tail is thus
\begin{align}\label{DLgtail}
\mathcal{D}L^\text{tail}_g &= -\frac{2G^2 M}{5 c^8}
\ln\left(\frac{\tau_0^\text{DR}}{\tau_0^\text{HR}}\right)
\,\left(I^{(3)}_{ij}\right)^2 \nonumber\\ &= \frac{G^2 m}{5 c^8}
\left[ - \frac{1}{\varepsilon} + \frac{41}{30} + 2
  \ln\left(\frac{\sqrt{\bar{q}}\,s_0}{\ell_0}\right)\right]
\left(I^{(3)}_{ij}\right)^2\,,
\end{align}
where we approximated $M=m+\mathcal{O}(1/c^2)$ in the second
equality. Thus, the pole in~\eqref{DLgtail} will indeed cancel out the
pole in the instantaneous part of the Fokker action [see
Eq.~\eqref{DLginst}].

\section{Determination of the ambiguity parameters}
\label{sec:amb}

We gather and recapitulate our results from the previous
sections. Recall that in Paper~I, the 4PN Fokker Lagrangian
constructed in harmonic coordinates initially depended on the
arbitrary constant parameter
\begin{equation}\label{alpha}
\alpha = \ln\left(\frac{r_0}{s_0}\right)\,,
\end{equation}
which was then adjusted to the value $\alpha=\frac{811}{672}$ by
comparison to the circular orbit limit of the binary's conserved
energy in the small mass ratio limit. We therefore have to:
\begin{enumerate}
\item Restore the arbitrariness of the parameter $\alpha$ by adding to
  the end result of Paper~I the contribution
\begin{equation}\label{DLgalpha}
\mathcal{D}L^\alpha_g = \frac{2G^2 m}{5 c^8} \left(\alpha -
\frac{811}{672}\right) \,\left(I^{(3)}_{ij}\right)^2 \,;
\end{equation}
\item Add the difference between the DR and HR evaluations of the IR
  divergences in the instantaneous part of the gravitational action,
  as computed using the method $n+2$, and whose result has been
  obtained in Eq.~\eqref{DLginst};
\item Subtract off the particular surface term given by
  Eq.~\eqref{DLgsurf} and which was necessary in the HR scheme for
  having a Lagrangian starting at the quadratic order with the
  propagator form $\propto h\Box h$;
\item Finally, add the difference between the radiation non-local
  tails in DR and HR as obtained in~\eqref{DLgtail}.
\end{enumerate}
Concerning the matter part $L_m$ of the Fokker Lagrangian, nothing is
to be changed with respect to the result of Paper~I since there are no
IR divergences therein and $L_m$ stands correct in DR. Finally our
full DR Lagrangian reads
\begin{equation}\label{Lcomplete}
L = L^\text{Paper~I} + \mathcal{D}L^\alpha_g +
\mathcal{D}L^\text{inst}_g - \mathcal{D}L^\text{surf}_g +
\mathcal{D}L^\text{tail}_g\,.
\end{equation}

Inserting our explicit results we find that the poles properly cancel
out as announced; furthermore the constants $r_0$, $s_0$ and $\ell_0$
also correctly disappear, and so does the irrational number $\bar{q} =
4\pi\,\ue^{\gamma_\text{E}}$. The modification of the Lagrangian then
takes exactly the form postulated in Eq.~(2.4) of Paper~II, namely
\begin{equation}\label{LDR}
L = L^\text{Paper~I} + \frac{G^4 m
  \,m_1^2m_2^2}{c^8r_{12}^4}\Bigl(\delta_1 (n_{12}v_{12})^2 + \delta_2
v_{12}^2 \Bigr)\,,
\end{equation}
but where the two ambiguity parameters $\delta_1$ and $\delta_2$ are
now unambiguously determined, as
\begin{equation}\label{ambparam}
\delta_1 = -\frac{2179}{315} \,,\qquad\delta_2
= \frac{192}{35} \,.
\end{equation}
This is exactly the values we obtained in Paper~II by demanding that the
conserved energy and periastron advance for circular orbits recover the GSF
calculations in the small mass-ratio limit. This result confirms the soundness
of the postulated form of the ambiguities in Paper~II and shows the power of
dimensional regularization for handling both UV and IR divergences in the
problem of motion in classical GR.

Remarkably, the value $\kappa=\frac{41}{60}$ we have obtained in our result for the tail [see Eq.~\eqref{StailDR1}], agrees with the result found by
Galley \textit{et al}~\cite{GLPR16} in their computation of the tail term in
$d$ dimensions (including both conservative and dissipative effects) by means
of EFT methods. This indicates that when the EFT calculation will be fully
completed at the 4PN order~\cite{FS4PN,FStail,GLPR16,FMSS16}, their result
will be free of any ambiguity like ours.

\acknowledgments 

It is a pleasure to thank Tanguy Marchand for having checked the
calculation of the tail effect in Sec.~\ref{sec:tail}. L.Bl. and G.F.
acknowledge a very useful and productive ``Workshop on 
analytical methods in General Relativity'' organized by Rafael Porto
and Riccardo Sturani at ICTP/SAIFR in S\~ao Paulo, Brazil.
L.Be. acknowledges financial support provided under the European
Union's H2020 ERC Consolidator Grant ``Matter and strong-field
gravity: New frontiers in Einstein's theory'' grant agreement
no. MaGRaTh646597.

\appendix

\section{Homogeneous solutions of the wave equation in $d+1$ dimensions} 
\label{app:hom}

The general ``monopolar'' homogeneous retarded solution of the wave
equation in $d+1$ dimensions (where $d=3+\varepsilon$), such that
$\Box \tilde{f}_\text{ret}(t,r) = 0$, reads, following the
notation~\eqref{Itilde},
\begin{align}\label{ftilde}
\tilde{f}_\text{ret}(r,t) &= - 4\pi \int_{-\infty}^{+\infty} \ud
t'\,G_\text{ret}(\mathbf{x},t-t') \,f(t')\nonumber\\ &=
\frac{\tilde{k}}{r^{d-2}}\int_1^{+\infty} \ud
y\,\gamma_{\frac{1-d}{2}}(y)\,f\Bigl(t-\frac{r y}{c}\Bigr)\,,
\end{align}
or, in more details, recalling
$\tilde{k}=\Gamma(\frac{d}{2}-1)/\pi^{\frac{d}{2}-1}$ and the function
$\gamma_s(y)$ displayed in Eq.~\eqref{gammas},
\begin{align}\label{detail}
\tilde{f}_\text{ret}(r,t) = \frac{2}{\pi^{\frac{\varepsilon}{2}}}
\frac{r^{-1-\varepsilon}}{\Gamma(-\frac{\varepsilon}{2})}\int_1^{+\infty}
\ud y\,(y^2-1)^{-1-\frac{\varepsilon}{2}}\,f\Bigl(t-\frac{r
  y}{c}\Bigr)\,.
\end{align}

In this Appendix we shall mostly investigate the post-Newtonian
expansion of that solution. We notice that by posing $\tau=r y/c$ we
are fixing the argument of the function $f$ in~\eqref{detail}, and
then the formal PN expansion $c\to +\infty$ becomes equivalent to a
formal expansion when $y\to +\infty$, which can simply be evaluated by
inserting into~\eqref{detail} the series
\begin{equation}\label{yinfty}
(y^2-1)^{-1-\frac{\varepsilon}{2}} = \sum_{k=0}^{+\infty}
  \frac{(-)^k}{k!}\frac{\Gamma(-\frac{\varepsilon}{2})}
       {\Gamma(-k-\frac{\varepsilon}{2})}\,y^{-2-2k-\varepsilon}\,.
\end{equation}
In this way we readily obtain
\begin{equation}\label{ftildeexp}
\tilde{f}_\text{ret} =
\frac{2}{\pi^{\frac{\varepsilon}{2}}c^{1+\varepsilon}}\sum_{k=0}^{+\infty}
\frac{(-)^k}{k!} \frac{(r/c)^{2k}}{\Gamma(-k-\frac{\varepsilon}{2})}
\int_{r/c}^{+\infty}\ud\tau\,\tau^{-2-2k-\varepsilon}\,f(t-\tau)\,.
\end{equation}
At this stage we split the integral according to
$\int_{r/c}^{+\infty}=-\int_0^{r/c}+\int_0^{+\infty}$. The two pieces
will respectively yield the decomposition of Eq.~\eqref{ftildeexp}
into ``even'' and ``odd'' pieces in the limit $\varepsilon\to 0$,
where we are following the standard PN terminology, \textit{i.e.},
meaning the parity of the power of $1/c$ in front. Thus,
\begin{equation}\label{evenodd}
\tilde{f}_\text{ret} =
\tilde{f}_\text{even}+\tilde{f}^\text{odd}_\text{ret}\,.
\end{equation}

In the even piece, corresponding to (minus) the integral from $0$ to
$r/c$, we are allowed to formally expand the integrand when $\tau\to
0$, since by definition $r/c\to 0$ for the PN expansion. At first
sight, this yields a complicated double infinite summation, but which
can be drastically simplified thanks to the formula
\begin{equation}\label{formule}
\sum_{k=0}^{+\infty} \frac{(-)^k}{k!}
\frac{1}{\left(k+\frac{1-p+\varepsilon}{2}\right)
  \Gamma(-k-\frac{\varepsilon}{2})} =
\frac{\Gamma(\frac{1-p+\varepsilon}{2})}{\Gamma(\frac{1-p}{2})}\,.
\end{equation}
Although it is valid for any $p\in\mathbb{N}$, this formula gives zero
whenever $p$ is an odd integer. Thus only will contribute the even
values $p=2j$, reflecting the even character, in the PN sense, of that
term. Furthermore we get a ``local'' expansion in any dimensions,
given by
\begin{equation}\label{even}
\tilde{f}_\text{even} =
\frac{r^{-1-\varepsilon}}{\pi^{\frac{1+\varepsilon}{2}}}
\sum_{j=0}^{+\infty} \frac{(-)^j}{2^{2j} j!}
\,\Gamma\bigl(\tfrac{1+\varepsilon}{2}-j\bigr)
\left(\frac{r}{c}\right)^{2j}\,f^{(2j)}(t) \,.
\end{equation}
As for the odd piece, corresponding to the integral from $0$ to
$+\infty$, it will irreducibly be given by a non-local integral
(``violation of Huygens' principle''), except when $\varepsilon=0$. We
perform a series of integrations by parts to arrive at an expression
which is manifestly finite in the limit $\varepsilon\to 0$:
\begin{equation}\label{odd}
\tilde{f}^\text{odd}_\text{ret} = -
\frac{1}{2\pi^{\frac{\varepsilon}{2}}c^{1+\varepsilon}}
\frac{\Gamma(\frac{1+\varepsilon}{2})}{\Gamma(1-\frac{\varepsilon}{2})}
\sum_{j=0}^{+\infty} \frac{1}{2^{2j}j!}
\frac{(r/c)^{2j}}{\Gamma(j+\frac{3+\varepsilon}{2})}
\,\int_0^{+\infty}\ud\tau\,\tau^{-\varepsilon}\,f^{(2j+2)}(t-\tau) \,.
\end{equation}
Notice that this expression, unlike~\eqref{even}, is regular when
$r\to 0$, \textit{i.e.}, $\tilde{f}^\text{odd}_\text{ret}\in
C^\infty(\mathbb{R})$. We straightforwardly check that
Eqs.~\eqref{even} and~\eqref{odd} recover in the limit $\varepsilon\to
0$ the usual even and odd parts of the PN expansion of the monopolar
wave (in particular, $\tilde{f}^\text{odd}_\text{ret}$ becomes local
in this limit):
\begin{subequations}\label{feps0}
\begin{align}
\tilde{f}_\text{ret}(r,t){\Big|}_{\varepsilon=0} &=
\frac{f(t-r/c)}{r}\,,\\ \tilde{f}_\text{even}(r,t){\Big|}_{\varepsilon=0}
&=
\sum_{j=0}^{+\infty}\frac{r^{2j-1}}{(2j)!c^{2j}}\,f^{(2j)}(t)\,\\
  \tilde{f}^\text{odd}_\text{ret}(r,t){\Big|}_{\varepsilon=0}
&= -
\sum_{j=0}^{+\infty}\frac{r^{2j}}{(2j+1)!c^{2j+1}}\,f^{(2j+1)}(t)\,.
\end{align}
\end{subequations}

The same analysis but done for the \textit{advanced} monopolar
homogeneous solution, \textit{i.e.}, using the advanced Green's
function [given by Eq.~\eqref{Gret} with $\theta(-t-r)$ in place of
  $\theta(t-r)$], gives
\begin{equation}\label{fadv}
\tilde{f}_\text{adv} =
\tilde{f}_\text{even}+\tilde{f}^\text{odd}_\text{adv}\,,
\end{equation}
where the even part is the same as before, and with the advanced odd
part
\begin{equation}\label{oddadv}
\tilde{f}^\text{odd}_\text{adv} = -
\frac{1}{2\pi^{\frac{\varepsilon}{2}}c^{1+\varepsilon}}
\frac{\Gamma(\frac{1+\varepsilon}{2})}{\Gamma(1-\frac{\varepsilon}{2})}
\sum_{j=0}^{+\infty} \frac{1}{2^{2j}j!}
\frac{(r/c)^{2j}}{\Gamma(j+\frac{3+\varepsilon}{2})}
\,\int_0^{+\infty}\ud\tau\,\tau^{-\varepsilon}\,f^{(2j+2)}(t+\tau) \,.
\end{equation}
In the limit $\varepsilon\to 0$ we evidently get
$\tilde{f}^\text{odd}_\text{adv}{\big|}_{\varepsilon=0} =
-\tilde{f}^\text{odd}_\text{ret}{\big|}_{\varepsilon=0}$. Further, we
define the associated symmetric and anti-symmetric solutions,
\begin{subequations}\label{symasym}
\begin{align}
\tilde{f}_\text{sym} &= \frac{1}{2}\bigl(\tilde{f}_\text{ret} +
\tilde{f}_\text{adv}\bigr) =
\tilde{f}_\text{even}+\frac{1}{2}\bigl(\tilde{f}^\text{odd}_\text{ret}
+ \tilde{f}^\text{odd}_\text{adv}\bigr)\,,\\ \tilde{f}_\text{asym} &=
\frac{1}{2}\bigl(\tilde{f}_\text{ret} - \tilde{f}_\text{adv}\bigr) =
\frac{1}{2}\bigl(\tilde{f}^\text{odd}_\text{ret} -
\tilde{f}^\text{odd}_\text{adv}\bigr)\,.
\end{align}
\end{subequations}
In particular, the anti-symmetric solution is non-local (except when
$\varepsilon = 0$), regular when $r\to 0$, and becomes purely odd in
the PN sense when $\varepsilon = 0$,
\begin{equation}\label{asym}
\tilde{f}_\text{asym} = -
\frac{1}{4\pi^{\frac{\varepsilon}{2}}c^{1+\varepsilon}}
\frac{\Gamma(\frac{1+\varepsilon}{2})}{\Gamma(1-\frac{\varepsilon}{2})}
\sum_{j=0}^{+\infty} \frac{1}{2^{2j}j!}
\frac{(r/c)^{2j}}{\Gamma(j+\frac{3+\varepsilon}{2})}
\,\int_0^{+\infty}\ud\tau\,\tau^{-\varepsilon}\Bigl[f^{(2j+2)}(t-\tau)
  - f^{(2j+2)}(t+\tau)\Bigr] \,.
\end{equation}

The most general ``multipolar'' homogeneous retarded solution will be
obtained by repeatedly applying spatial differentiations on the latter
monopolar solution, hence
\begin{equation}\label{multipolar}
\tilde{H}_\text{ret}(\mathbf{x},t) = \sum_{\ell=0}^{+\infty}
\,\hat{\partial}_L\tilde{f}_\text{ret}^L(r,t)\,,
\end{equation}
where $\hat{\partial}_L$ denotes the STF product of $\ell$ spatial
derivatives (and $L=i_1\cdots i_\ell$). Similarly one can define the
advanced, symmetric and anti-symmetric multipolar solutions. For
instance, the anti-symmetric solution can be re-written in the
manifestly regular form
\begin{align}\label{multanti}
\tilde{H}_\text{asym} &= - \frac{1}{4\pi^{\frac{\varepsilon}{2}}}
\frac{\Gamma(\frac{1+\varepsilon}{2})}
     {\Gamma(1-\frac{\varepsilon}{2})}\sum_{\ell=0}^{+\infty} \frac{1}
     {2^{\ell} \Gamma(\ell+\frac{3+\varepsilon}{2})}
     \sum_{j=0}^{+\infty}\frac{\Delta^{-j}
       \hat{x}_L}{c^{2j+2\ell+1+\varepsilon}}\nonumber\\ &\qquad\qquad\times
     \int_0^{+\infty}\ud\tau\,\tau^{-\varepsilon}
     \Bigl[f_L^{(2j+2\ell+2)}(t-\tau) -
       f_L^{(2j+2\ell+2)}(t+\tau)\Bigr]\,,
\end{align}
where we recall the short-hand notation
\begin{equation}\label{Deltaj2}
\Delta^{-j} \hat{x}_L = \frac{\Gamma(\ell+\frac{3+\varepsilon}{2})}
      {\Gamma(\ell+j+\frac{3+\varepsilon}{2})} \,\frac{r^{2j}
        \hat{x}_L}{2^{2j} j!}\,.
\end{equation}
The homogeneous solution investigated in Sec.~\ref{sec:NZ}, and that
we computed directly from a near-zone expansion, is precisely of the
previous anti-symmetric type~\eqref{multanti}. We showed this by going
to the Fourier domain [see Eqs.~\eqref{solvefF} and~\eqref{Fourier}].

\section{Multipole expansion of elementary functions in $d$ 
dimensions}
\label{app:mult}

For our computation of the difference between the DR and HR
prescriptions for the IR regularization of integrals at infinity in
Sec.~\ref{sec:diffIR}, we need to control the expansion at infinity
($r\to+\infty$) of non-linear potentials in $d$ dimensions. These
potentials are defined by means of elementary solutions of the Poisson
or d'Alembert equation in $d$ dimensions, the simplest one being the
famous Fock kernel obeying in $d$ dimensions
\begin{equation}\label{g}
\Delta g = r_1^{2-d}r_2^{2-d}\,.
\end{equation}
The exact expression in $3$ dimensions is $g^{(\varepsilon=0)} =
\ln(r_1+r_2+r_{12})$~\cite{Fock}. The explicit form of the solution in
$d$ dimensions has been obtained in the Appendix~C of
Ref.~\cite{BDE04}. In the Appendix~B of Paper~I we have given the
local expansion of that function in $d$ dimensions near the
singularities (when $r_{1}$ or $r_{2}\to 0$). Here we compute the far
zone expansion when $r\to+\infty$, that we shall refer to as a
multipole expansion denoted $\mathcal{M}(g)$.

Suppose we want to compute the multipole expansion $\mathcal{M}(P)$ of
some elementary potential $P$, solution of the wave equation $\Box P =
\sigma$, where $\sigma$ is some source term with non compact support
like in~\eqref{g}. In the usual post-Newtonian (or near zone)
iteration scheme, neglecting time-odd contributions, the potential is
given by $P = \mathcal{I}^{-1}\sigma$ where the usual symmetric
propagator reads
\begin{align}\label{sympropag}
\mathcal{I}^{-1} = \sum_{p=0}^{+\infty}
\left(\frac{1}{c}\frac{\partial}{\partial t}\right)^p \Delta^{-p-1}\,.
\end{align}
Now the far-zone expansion $\mathcal{M}(P)$ will be obtained from the
far-zone expansion $\mathcal{M}(\sigma)$ of the corresponding source
term by application of~\eqref{sympropag}, but for a non-compact
support source it is known that there is also a homogeneous solution
of the \textit{symmetric} type to be added, and which is specified by
Eq.~(3.23) of Ref.~\cite{PB02}. Generalizing the formula to $d$
dimensions, this means that the solution is the sum of a particular
solution obtained by application of Eq.~\eqref{sympropag}, plus a
specific homogeneous symmetric one,
\begin{equation}\label{matcheq}
\mathcal{M}(P) = \mathcal{I}^{-1}[\mathcal{M}(\sigma)] -
\frac{1}{4\pi} \sum_{\ell=0}^{+\infty} \frac{(-)^\ell}{\ell!}
\,\overline{\partial_L \tilde{\sigma}_\text{sym}^L} \,,
\end{equation}
where the overbar on the homogeneous solution means the PN or
near-zone expansion, and, following the Appendix~\ref{app:hom}, the
homogeneous symmetric solution reads
\begin{equation}\label{sigmasym}
\tilde{\sigma}_\text{sym}^L(r,t) =
\frac{\tilde{k}}{r^{d-2}}\int_1^{+\infty} \ud
y\,\gamma_{\frac{1-d}{2}}(y)\Bigl[\sigma_L\bigl(t-r y/c\bigr) +
  \sigma_L\bigl(t+r y/c\bigr)\Bigr]\,.
\end{equation}
Here $\sigma_L$ denotes the $\ell$-th multipole moment of the source
$\sigma$ given (in non-STF guise) by
\begin{equation}\label{nonSTFmoment}
\sigma_L(t) = \int \ud^d\mathbf{x}'\,{x'}_L\,\sigma(\mathbf{x}',t)\,.
\end{equation}
Note that we are performing a full DR calculation, so the multipole
moment $\sigma_L$ is defined without invoking a finite part
regularization (based on some regulator $(r/r_0)^B$ with
$B\in\mathbb{C}$); instead, DR is taking care of the IR divergences,
appearing here due to the fact that the source $\sigma$ has a
non-compact spatial support. Similarly, the particular solution or
first term in Eq.~\eqref{matcheq}, is defined in a pure DR way, with
the iterated Poisson operator $\Delta^{-p-1}$ in~\eqref{sympropag}
acting on each term of the multipole expansion of the source
$\mathcal{M}(\sigma)$, whose general structure in $d$ dimensions is
provided by Eq.~\eqref{Fdevddim}. It is clear that the Poisson
operator and its iterated version make sense when applied to such
terms [see, \textit{e.g.},~\eqref{Deltaj2}].

Finally, because of the overbar prescription in Eq.~\eqref{matcheq},
we need the post-Newtonian or near-zone expansion of the object
$\tilde{\sigma}_\text{sym}^L$. The PN expansion of the homogeneous
symmetric solution has been investigated in the previous
App.~\ref{app:hom}. It consists essentially of even contributions but
also, in $d$ dimensions, or some residual non-local odd terms [see
Eqs.~\eqref{symasym}]. The odd terms will disappear in $3$ dimensions;
we neglect these since they are dissipative contributions. Thus we
simply assimilate the symmetric part with the even part, and we get,
from Eq.~\eqref{even},
\begin{equation}\label{sigmaeven}
\tilde{\sigma}^L_\text{sym} =
\frac{r^{-1-\varepsilon}}{\pi^{\frac{1+\varepsilon}{2}}}
\sum_{j=0}^{+\infty} \frac{(-)^j}{2^{2j} j!}
\,\Gamma\bigl(\tfrac{1+\varepsilon}{2}-j\bigr)
\left(\frac{r}{c}\right)^{2j}\,\sigma_L^{(2j)}(t) \,.
\end{equation}

We have applied the previous formulas to the source term $\sigma =
r_1^{2-d}r_2^{2-d}$ in Eq.~\eqref{g}. Defining $g$ and $f$ such that,
up to the 1PN order,
\begin{equation}\label{Pgf}
P = g + \frac{1}{2c^2}\,\partial_t^2f +
\mathcal{O}\left(\frac{1}{c^4}\right)\,,
\end{equation}
we have $\Delta g = \sigma$ and $\Delta f = 2g$ in this convention.
We obtain\footnote{With our notation for multi-indices meaning, for
  instance,
\begin{align*}
\hat{n}^{M-2S} &= \text{STF}[n^{i_1}\cdots n^{i_{m-2s}}]\,,
\\ \hat{n}^{M-2S} \hat{y}_1^{L-S,S'} \hat{y}_2^{M-LS,S'} &=
\hat{n}^{i_1\cdots i_{m-2s}} \hat{y}_1^{i_1\cdots i_{\ell-s}j_1\cdots
  j_s} \hat{y}_2^{i_{\ell-s+1}\cdots i_{m-2s}j_1\cdots j_s} \,.
\end{align*}}
\begin{subequations}
\begin{align}
\mathcal{M}(g) &= \frac{r_{12}^{1-\varepsilon}}{1-\varepsilon}
\sum_{\ell=0}^{+\infty} \frac{2^{\,\ell-1}}{(\ell+1)!}
\frac{\Gamma(\ell+\frac{\varepsilon+1}{2})}
     {\Gamma(\frac{\varepsilon+1}{2})}
     \frac{\hat{n}^L}{r^{\ell+1+\varepsilon}} \sum_{s=0}^\ell
     y_1^{\langle L-S} y_2^{S\rangle} \nonumber\\ & +
     \frac{1}{\bigl[\Gamma(\frac{1+\varepsilon}{2})\bigr]^2}
     \sum_{m=0}^{+\infty} \frac{2^{\,m-2}}{r^{m+2\varepsilon}}
     \sum_{s=0}^{[\frac{m}{2}]}
     \frac{\Gamma(\frac{3+\varepsilon}{2}+m-2s)}
          {\Gamma(\frac{3+\varepsilon}{2}+m-s)}
          \frac{\hat{n}^{M-2S}}{(m-s+\varepsilon)
            (s+\frac{\varepsilon-1}{2})(2s)!!}  \nonumber\\ & \qquad
          \quad \times \sum_{\ell=0}^m \frac{\Gamma(\ell +
            \frac{\varepsilon+1}{2})}{(\ell-s)!}
          \frac{\Gamma(m-\ell+\frac{\varepsilon+1}{2})}{(m-\ell-s)!}
          \hat{y}_1^{L-S,S'} \hat{y}_2^{M-LS,S'}\,,\\
\mathcal{M}(f) &= \frac{r_{12}^{1-\varepsilon}}{(1-\varepsilon)^2}
\sum_{\ell=0}^{+\infty} \frac{2^{\,\ell-1}}{(\ell+1)!}
\frac{\Gamma(\ell+\frac{\varepsilon-1}{2})}
     {\Gamma(\frac{\varepsilon-1}{2})} \bigg[\sum_{s=0}^\ell
       y_1^{\langle L-S} y_2^{S\rangle} \bigg( r^2
       -\frac{(2\ell+\varepsilon-1)}{(2\ell+\varepsilon+3)(\ell+2)}
       \nonumber\\ & \qquad \quad \times \Big( y_1^2(\ell-s+1)+y_2^2
       (s+1)-\frac{2r_{12}^2}{3-\varepsilon} (\ell-s+1) (s+1) \Big)
       \bigg) \bigg]
     \frac{\hat{n}^L}{r^{\ell+1+\varepsilon}}\nonumber\\ & +
     \frac{1}{\bigl[\Gamma(\frac{1+\varepsilon}{2})\bigr]^2}
     \sum_{m=0}^{+\infty} \frac{2^{\,m-3}}{r^{m-2+2\varepsilon}}
     \sum_{s=0}^{[\frac{m}{2}]}
     \frac{\Gamma(\frac{3+\varepsilon}{2}+m-2s)
       \Gamma(m-s-1+\varepsilon)\Gamma(s+\frac{\varepsilon-3}{2})}
          {\Gamma(\frac{3+\varepsilon}{2}+m-s)
            \Gamma(m-s+1+\varepsilon)\Gamma(s+\frac{\varepsilon+1}{2})}
          \frac{\hat{n}^{M-2S}}{(2s)!!} \nonumber\\ & \qquad \quad
          \times \sum_{\ell=0}^m \frac{\Gamma(\ell +
            \frac{\varepsilon+1}{2})}{(\ell-s)!}
          \frac{\Gamma(m-\ell+\frac{\varepsilon+1}{2})}{(m-\ell-s)!}
          \hat{y}_1^{L-S,S'} \hat{y}_2^{M-LS,S'} \,.
\end{align}
\end{subequations}
Similarly, in our calculations we have also to consider the potentials
$f_{12}$ and $f_{21}$ obeying
\begin{equation}
\Delta f_{12} = r_1^{4-d}r_2^{2-d}\,,\qquad \Delta f_{21} =
r_1^{2-d}r_2^{4-d}\,,
\end{equation}
and we obtain, for instance,
\begin{align}
\mathcal{M}(f_{12}) &= \frac{r_{12}^{3-\varepsilon}}{3-\varepsilon}
\sum_{\ell=0}^{+\infty} \frac{2^{\,\ell-1}}{(\ell+2)!}
\frac{\Gamma(\ell+\frac{\varepsilon+1}{2})}
     {\Gamma(\frac{\varepsilon+1}{2})}
     \frac{\hat{n}^L}{r^{\ell+1+\varepsilon}}\sum_{s=0}^\ell (s+1)
     y_1^{\langle L-S} y_2^{S\rangle} \nonumber\\ & -
     \frac{1}{(1-\varepsilon)
       \bigl[\Gamma(\frac{\varepsilon-1}{2})\bigr]^2}
     \sum_{m=0}^{+\infty} \frac{2^{\,m-1}}{r^{m+2\varepsilon}}
     \sum_{s=0}^{[\frac{m}{2}]}
     \frac{\Gamma(\frac{3+\varepsilon}{2}+m-2s)}
          {\Gamma(\frac{3+\varepsilon}{2}+m-s)}
          \frac{\hat{n}^{M-2S}}{(2s)!!} \nonumber\\ & \qquad \quad
          \times \sum_{\ell=0}^m \frac{\Gamma(\ell +
            \frac{\varepsilon-1}{2})}{(\ell-s)!}
          \frac{\Gamma(m-\ell+\frac{\varepsilon+1}{2})}{(m-\ell-s)!}
          \hat{y}_1^{L-S,S'} \hat{y}_2^{M-LS,S'} \nonumber\\ & \qquad
          \quad \times \bigg[
            \frac{r^2}{(m-s-1+\varepsilon)(s+\frac{\varepsilon-3}{2})}
            - \frac{(2\ell+\varepsilon-1)y_1^2}
            {(2\ell+\varepsilon+3)(m-s+\varepsilon)
              (s+\frac{\varepsilon-1}{2})} \bigg]\,.
\end{align}
The above formulas have been extensively used to control the IR
divergences in the gravitational part of the Fokker action in
Sec.~\eqref{sec:diffIR}. However we have found that in fact, the
result of our computation of the difference DR$-$HR does not depend on
the detailed prescription we followed to control the homogeneous
anti-symmetric solution in Eq.~\eqref{matcheq}. The independence with
respect to the added homogeneous solution in Eq.~\eqref{matcheq} is
certainly a good sign of the solidness of our result.

\section{Distributional limits of the function $\gamma_s(z)$}
\label{app:distr}

The function $\gamma_s(z)$ defined by Eq.~\eqref{gammas} is zero in an
ordinary sense for strictly negative integer values $s=-1-\ell$ (where
$\ell\in\mathbb{N}$). In this Appendix we compute
$\gamma_{-1-\ell}(z)$ in the sense of distributions. From
Eq.~\eqref{gammas} we have
\begin{equation}\label{gammadistr}
\gamma_{-1-\ell-\frac{\varepsilon}{2}}(z) =
\frac{2\sqrt{\pi}}{\Gamma(-\ell-\frac{\varepsilon}{2})
  \Gamma(\ell+\frac{1+\varepsilon}{2})}
\,\big(z^2-1\bigr)^{-1-\ell-\frac{\varepsilon}{2}}\theta(z-1)\,.
\end{equation}
We added the Heaviside step function $\theta(z-1)$ to recall that this
expression is defined only for $z>1$. Considered as a distribution
(indexed by a parameter $\varepsilon\in\mathbb{C}$),
Eq.~\eqref{gammadistr} is to be applied on test functions $\varphi(z)$
that are at once smooth, \textit{i.e.}, $\varphi\in
C^\infty(\mathbb{R})$, and with compact support. Hence,
\begin{equation}\label{distrtest1}
\langle\gamma_{-1-\ell-\frac{\varepsilon}{2}}, \varphi\rangle =
\frac{2\sqrt{\pi}}{\Gamma(-\ell-\frac{\varepsilon}{2})
  \Gamma(\ell+\frac{1+\varepsilon}{2})} \int_1^{+\infty}\ud
z\,\big(z^2-1\bigr)^{-1-\ell-\frac{\varepsilon}{2}}\varphi(z)\,.
\end{equation}
Under this form we see that the limit $\varepsilon\to 0$ is
ill-defined at the bound $z=1$, but can made finite by performing some
integrations by parts. The surface terms will always be zero by
analytic continuation in $\varepsilon$ at the bound $z=1$, and because
the test function has a compact support. After $\ell+1$ integrations
by parts we obtain
\begin{equation}\label{distrtest2}
\langle\gamma_{-1-\ell-\frac{\varepsilon}{2}}, \varphi\rangle =
(-)^{\ell+1}\frac{2\sqrt{\pi}}{\Gamma(1-\frac{\varepsilon}{2})
  \Gamma(\ell+\frac{1+\varepsilon}{2})} \int_1^{+\infty}\ud
z\,\big(z-1\bigr)^{-\frac{\varepsilon}{2}} \left(\frac{\ud}{\ud
  z}\right)^{\ell+1}
\!\left[\big(z+1\bigr)^{-1-\ell-\frac{\varepsilon}{2}}\varphi(z)\right]\,,
\end{equation}
and, under that form, we can directly take the limit $\varepsilon\to
0$ with result
\begin{equation}\label{distrtest3}
\langle\gamma_{-1-\ell}, \varphi\rangle =
\frac{(-)^{\ell}2^{\ell+1}}{(2\ell-1)!!}\left(\frac{\ud}{\ud
  z}\right)^{\ell}
\!\!\biggl[\frac{\varphi(z)}{\big(z+1\bigr)^{\ell+1}}\biggr]{\bigg|}_{z=1}
\,.
\end{equation}
More explicitly this gives
\begin{subequations}\label{distrtest4}
\begin{align}
\langle\gamma_{-1-\ell}, \varphi\rangle &= \sum_{i=0}^{\ell} (-)^i
\alpha_i^\ell\,\varphi^{(i)}(1) \,,\\ \text{where}\quad \alpha_i^\ell
&= \frac{2^{i-\ell}}{(2\ell-1)!!}\frac{(2\ell-i)!}{i! (\ell-i)!} \,.
\end{align}
\end{subequations}
So, finally the result for $\gamma_{-1-\ell}$ when viewed as a
distribution reads
\begin{align}\label{distr}
\gamma_{-1-\ell}(z) = \sum_{i=0}^{\ell}
\alpha_i^\ell\,\delta^{(i)}(z-1) \,,
\end{align}
with $\delta^{(i)}$ being the $i$-th derivative of the Dirac
function. In particular $\gamma_{-1}(z)=\delta(z-1)$ recovers the fact
that the Green's function~\eqref{Gret} reduces in $3+1$ dimensions to
the usual
\begin{align}\label{G3d}
G_\text{ret}^{(\varepsilon=0)}(\mathbf{x},t) = -
\frac{\delta(t-r)}{4\pi\,r}\,.
\end{align}

\section{Computation of the coefficients $C_\ell^{p,k}$}
\label{app:coeffC}

These coefficients, defined in $d=3+\varepsilon$ dimensions by
Eq.~\eqref{coeffC}, are written in the form
\begin{equation}\label{Celleps}
C_\ell^{p,k} = \frac{4\pi}{\Gamma(-\frac{\varepsilon}{2})
  \Gamma(-\ell-\frac{\varepsilon}{2})
  \Gamma(\frac{1+\varepsilon}{2})\Gamma(\ell+\frac{1+\varepsilon}{2})}
\,L^p_{a,b,c}\,,
\end{equation}
together with the following definition of the double integral,
\begin{equation}\label{Lpabc}
L^p_{a,b,c} = \int_1^{+\infty} \!\ud y
\,y^p\,(y^2-1)^a\int_1^{+\infty} \!\ud z \,(z^2-1)^b\,(y+z)^c\,,
\end{equation}
and the particular set of coefficients $a=-1-\frac{\varepsilon}{2}$,
$b=-\ell-1-\frac{\varepsilon}{2}$, and $c=\ell+k-2+2\varepsilon-\eta$, where
the parameter $\eta$ was introduced in Eq.~\eqref{phiFP}.

The integral~\eqref{Lpabc} is computed by first relating it to the
simpler integral corresponding to $p=0$, namely $K_{a,b,c} =
L^0_{a,b,c}$ or
\begin{equation}\label{Kabc}
K_{a,b,c} = \int_1^{+\infty} \!\ud y \,(y^2-1)^a\int_1^{+\infty} \!\ud
z \,(z^2-1)^b\,(y+z)^c\,.
\end{equation}
The latter integral in turn converges for $\Re(a) > -1$,
$\Re(b) > -1$, $\Re(2a+c) < -1$, $\Re(2b+c) < -1$ and
$\Re(2a+2b+c) < -2$. Moreover, it admits an explicit closed-form
expression in terms of Eulerian $\Gamma$-functions,
\begin{equation}\label{formKabc}
K_{a,b,c} = \frac{\Gamma(a+1) \Gamma(b+1)
  \Gamma(-a-\frac{c}{2}-\frac{1}{2})
  \Gamma(-b-\frac{c}{2}-\frac{1}{2})
  \Gamma(-a-b-\frac{c}{2}-1)}{4\sqrt{\pi}
  \,\Gamma(-\frac{c}{2}+\frac{1}{2}) \Gamma(-a-b-c-1)}\,,
\end{equation}
so that, regarded as a function of $a$, $b$ and $c$, it can be
extended to the complex plane by analytic continuation, except for a
countable number of isolated points.

Finally it is very easy to relate $L^p_{a,b,c}$ to $K_{a,b,c}$. When
$p=2q$ is an even integer, we have
\begin{equation}\label{Leven}
L^{2q}_{a,b,c} = \sum_{i=0}^q \genfrac{(}{)}{0pt}{}{q}{i}
\,K_{a+i,b,c} \,,
\end{equation}
where $\genfrac{(}{)}{0pt}{}{q}{i}$ is the usual binomial
coefficient. And, when $\ell=2q+1$ is an odd integer, we go back to
the even case~\eqref{Leven} thanks to the formula
\begin{equation}\label{Lodd}
L^{2q+1}_{a,b,c} = \frac{1}{2}\Bigl[L^{2q}_{a+1,b,c-1} -
  L^{2q}_{a,b+1,c-1} + L^{2q}_{a,b,c+1}\Bigr] \,.
\end{equation}

With those formulas we can compute the $C_\ell^{p,k}$ for all required
values of $\ell$, $p$ and $k$. Note that there are some combinations
of $a$, $b$ and $c$ for which the $\varepsilon$'s disappear. In these
cases it is crucial to keep the parameter $\eta$ finite, and to
compute the expansion series when $\eta\to 0$ (for any $\varepsilon$,
\textit{i.e.}, before applying the limit $\varepsilon\to 0$). We find
that many individual terms behave like $1/\eta$ and are thus
ill-defined, but that these divergences always cancel out from the sum
of all these terms. Thus, at the end we always get a finite result
when $\eta=0$, which can then be evaluated in the limit
$\varepsilon\to 0$.

\bibliography{ListeRef_BBBFMc}

\end{document}